\documentclass[12pt, a4paper]{article}
\usepackage{amssymb, amsmath, bm, float, epsfig, color, slashed, subcaption, multirow, arydshln, mathtools, bbm}
\usepackage[centering]{geometry}
\usepackage[all,cmtip]{xy}
\usepackage{mathrsfs}

\makeatletter
\@addtoreset{equation}{section}
\makeatother

\def \m {\mathcal}

\def \t {\tilde}
\def \g {\gamma}

\begin{document}

\thispagestyle{empty}

\begin{flushright}
April 2021 \\
\end{flushright}
\vspace{30pt}
\begin{center}
{\Large\bf Pati-Salam model in curved space-time and sheaf quantization } \\

\vspace{47.5pt}
{\large De-Sheng Li\textsuperscript{1}{\footnote{lideshengjy@126.com}}} \\

\vspace{12.5pt}
{\it \textsuperscript{1}College of Electrical and Information Engineering , Hunan Institute of Engineering, Xiangtan 411100, P.R.China\\
\it \textsuperscript{2}  College of Physics, Mechanical and Electrical Engineering, Jishou University, Jishou 416000, P.R.China}


\end{center}

\vspace{25pt}
\begin{abstract}
\noindent
This note aims to give a self-contain and detail explanation about $U(k^{\prime})\times U(k)$ Pati-Salam model in curved space-time which derived from $(1+n)$-dimensional square root Lorentz manifold by self-parallel transportation principle. 
The concepts foundation of manifold from view point of category theory, fiber bundle and sheaf theories are reviewed.
There are extra $U(k^{\prime})\times U(k)$-principal bundle and $U(k)$-associated bundle than $(1+n)$-dimensional Lorentz manifold. The conservative currents on square root Lorentz manifold are discussed preliminary. A detail proof of relation from sheaf quantization to path integral quantization is given.

\end{abstract}

\newpage
\setcounter{page}{2}
\setcounter{footnote}{0}

\tableofcontents

\section{Introduction}
Square root metric manifold has extra $U(k^{\prime})\times U(k)$ principal bundle and $U(k)$-associated bundle than usual Lorentz manifold. These extra bundles gives us opportunity to construct Yang-Mills theory in curved space-time \cite{Li:2014fxl}, especially the Pati-Salam type Yang-Mills theory \cite{Pati:1974yy, Li:2020zij} in curved space-time. Sheaf as a natural mathematic structure being found by mathematicians \cite{kashiwara2013sheaves}, for example, Jean Leray, long ago. Sheaf theory has deep relation with fiber bundle theory \cite{husemoller1966fibre} (Yang-Mills theory \cite{Yang1954, schwartz2014quantum}) and superposition principle. Sheaf can be derived from contravariant functor in category theory, the sheaf cohomology and spectral sequences is fascinating and useful. The micro support language of sheaf theory \cite{kashiwara2013sheaves} from Mikio Sato might be popular in future mathematic- physicists. Sheaf as a basic language of topos from Grothendieck \cite{dieudonne1971elements}, ``we cannot even define a scheme without using scheaves'' \cite{hartshorne2013algebraic}. Sheaf quantization might be a method to quantize quantum field theory in curved space-time which avoiding problem of infinities \cite{Li:2014fxl, Isham:1999kb,Froyshov:2002qaq,Fukaya:2011ad,Bochicchio:2013aha,Cho:2013xda,Nakayama:2014tga,Viterbo2019, Kuwagaki:2022rpj}. The sheaf space is linear space and coherent with superposition principle, even the base manifold is curved. The sheaf quantization method is consistent with path integral quantization method. 

In this paper, the section 2 gives us a preliminary concepts introduction of category, functor; and the topological space, sheaf, manifold, bundle from the category point of view. The section 3 talk about Einstein-Cartan geometry of Lorentz and Riemann manifold. The section 4 is a brief introduction of generators of Clifford algebra. The section 5 describes the geometry framework of square root Lorentz manifold. Based on square root Lorentz manifold, the Pati-Salam model in curved space-time and Einstein-Cartan gravity are constructed. The section 6 discusses the formulation of sheaf quantization, and the relation between sheaf quantization and path integral.





\section{Category, functor, topological space, sheaf, manifold and fiber bundle}
\subsection{Category}
The category $\mathbf{C}$ consist of 
\begin{itemize}
\item a class $ob(\mathbf{C})$ of objects, for example, $a,b,c,d\in ob(\mathbf{C})$.
\item a class $hom(\mathbf{C})$ of morphisms, or arrows, or maps. 
\begin{eqnarray}
hom_{\mathbf{C}}(a,b)
\end{eqnarray}
represent all morphisms from $a$ to $b$ in category $\mathbf{C}$. For example, 
\begin{eqnarray}
f,g\in hom(a,b), \quad h\in hom(b,c), \quad i\in hom(a,c).
\end{eqnarray}
\item Composition of morphisms is, for objects $a,b,c\in ob(\mathbf{C})$,
\begin{eqnarray}
hom_{\mathbf{C}}(a,b)\times hom_{\mathbf{C}}(b,c)\rightarrow hom_{\mathbf{C}}(a,c).
\end{eqnarray}
The morphisms $hom(\mathbf{C})$ in category $\mathbf{C}$ satisfy the axiom of associativity and identity: 
\item (Associativity axiom) if 
\begin{eqnarray}
f:a\rightarrow b,\quad g:b\rightarrow c,\quad h:c\rightarrow d,
\end{eqnarray}
then
\begin{eqnarray}
h(gf)=(hg)f.
\end{eqnarray}
\item (Identity axiom) For every object $x,y\in ob(\mathbf{C})$, there exists a morphism 
\begin{eqnarray}
1_{x}:x\rightarrow x.
\end{eqnarray}
For every morphism $f\in hom(\mathbf{C})$
\begin{eqnarray}
f: x\rightarrow y,
\end{eqnarray}
we have
\begin{eqnarray}
1_{x} f=f=f 1_{y}.
\end{eqnarray}
\end{itemize}
\subsection{Functor}
Functors are structure-preserving maps between categories. A covariant functor $F$ from a category $\mathbf{C}$ to a category $\mathbf{D}$ is written
\begin{eqnarray}
F:\mathbf{C}\rightarrow \mathbf{D},
\end{eqnarray}
and the structure-preserving means
\begin{itemize}
\item for object $x\in ob(\mathbf{C})$ and $F(x)\in ob(\mathbf{D})$ and morphisms $f\in hom(\mathbf{C})$
\begin{eqnarray}
f:x\rightarrow y ,\quad F(f): F(x)\rightarrow F(y),
\end{eqnarray}
where
\begin{eqnarray}
f\in hom(\mathbf{C}), \quad F(f) \in hom(\mathbf{D}).
\end{eqnarray}
such that,
\item For every object $x\in ob(\mathbf{C})$,
\begin{eqnarray}
F(1_{x})=1_{F(x)};
\end{eqnarray}
\item for objects $x,y,z\in ob(\mathbf{C})$, all morphisms in $\mathbf{C}$
\begin{eqnarray}
f:x\rightarrow y, &\quad & g:y\rightarrow z,
\end{eqnarray}
the functor preserves the composition of morphisms
\begin{eqnarray}
F(gf)=F(g)F(f).
\end{eqnarray}
\end{itemize}
A contravariant functor like structure-preserving covariant functor from categories $\mathbf{C}$ to $\mathbf{D}$, but for morphism $f,g\in hom(\mathbf{C})$
\begin{eqnarray}
f:x\rightarrow y, & \Rightarrow &F(f): F(y)\rightarrow F(x),\\
F(gf)&=&F(f)F(g).
\end{eqnarray}
\subsection{Topological space}
The point $x_{0}\in x $ and the neighborhood of $x_{0}$ (an open covering )
\begin{eqnarray}
U_{x_{0}}=\{x|x\rightarrow x_{0} \}
\end{eqnarray}
can glun to topological space $\mathbf{X}$ (more precisely, an open covering Hausdorff space $\mathbf{X}$ )
\begin{eqnarray}
\mathbf{X}=\cup U_{x},
\end{eqnarray}
where 
\begin{eqnarray}
x\rightarrow x_{0}
\end{eqnarray}
is the direct limit from $x$ to $x_{0}$. For any point $x_{0}\in x$, there is open covering partial ordered set on topological space $\mathbf{X}$
\begin{eqnarray}
U_{x_{0}}\subset U^{1}_{x_{0}} \subset U^{2}_{x_{0}} \subset \cdots \mathbf{X}.
\end{eqnarray}
\subsection{Category viewing of topological space}
\begin{itemize}
\item The Topological space $\mathbf{X}$ is a category with objects
\begin{eqnarray}
U_{x_{0}}\in ob(\mathbf{X}),&\quad & x_{0} \in x.
\end{eqnarray}
and morphisms
\begin{eqnarray}
\subset, \cup,\cap \in hom(\mathbf{X}).
\end{eqnarray}
\item The category $\mathbf{Top}$ with objects
\begin{eqnarray}
\mathbf{X}\in ob(\mathbf{Top}).
\end{eqnarray}
and morphisms
\begin{eqnarray}
continuous\ \ map \in hom(\mathbf{Top}).
\end{eqnarray}
\end{itemize}
\subsection{Presheaf and sheaf}
$F(U_{x_{0}})$ is the presheaf on $U_{x_{0}}$ which is isomorphic to Abel group $\mathbf{A}$ ($F(U_{x_{0}})$ means all possible functions on neighborhood $U_{x_{0}}$)
\begin{eqnarray}
F:U_{x_{0}}\rightarrow F(U_{x_{0}}).
\end{eqnarray}
$F$ is a functor from neighborhood $U_{x_{0}}$ to presheaf $F(U_{x_{0}})$.
From presheaf $F(U_{x_{0}})$ to construct sheaf $F(\mathbf{X})$ satisfy the locality axiom and gluing axiom.
\begin{itemize}
\item (Locality axiom) If $U_{x_{0}}$ is an open covering of an open set $\mathbf{X}$, and if sections
\begin{eqnarray}
s_{x}, t_{x}\in F(\mathbf{X}),
\end{eqnarray}
such that for any $x_{0}\in \mathbf{X}$
\begin{eqnarray}
s|_{U_{x_{0}}}= t|_{U_{x_{0}}},
\end{eqnarray}
then
\begin{eqnarray}
s_{x}= t_{x},
\end{eqnarray}
where $s|_{U_{x_{0}}}$ is the section restricted to neighborhood of $x_{0}$.
\item (Gluing axiom) If
\begin{eqnarray}
x_{0},x_{1}\in \mathbf{X},
\end{eqnarray}
$U_{x_{0}}$ and $U_{x_{1}}$ are open covering of an open set $\mathbf{X}$, and for sections
\begin{eqnarray}
s_{x_{0}}\in F(U_{x_{0}}), &\quad & s_{x_{1}}\in F(U_{x_{1}}),
\end{eqnarray}
the sections agree on the overlap 
\begin{eqnarray}\label{gluon}
s_{x_{0}}|_{U_{x_{0}} \cap U_{x_{1}}}=s_{x_{1}}|_{U_{x_{1}} \cap U_{x_{0}}},
\end{eqnarray}
the presheaf gluing axiom (\ref{gluon}) can be presented by commutative diagram
\begin{eqnarray}
\xymatrix{
F(U_{x_{0}}) \ar[rr]^{\cap F(U_{x_{1}})} & & F(U_{x_{0}}\cap U_{x_{1}})\\
&&\\
U_{x_{0}}\ar[rr]^{\cap U_{x_{1}}} \ar[uu]_{F}& & U_{x_{0}}\cap U_{x_{1}} \ar[uu]_{F}
}
\end{eqnarray}
then there is a global section 
\begin{eqnarray}
s_{x}\in F(\mathbf{X}), \quad x\in \mathbf{X},
\end{eqnarray}
such that
\begin{eqnarray}
s_{x_{0}}= s_{x}|_{U_{x_{0}}}.
\end{eqnarray}
The stalk of $x_{0}$ is the sheaf space restricted to $x_{0}$
\begin{eqnarray}
F_{x_{0}}=F(\mathbf{X})|_{U_{x_{0}}}
=F(U_{x_{0}})/\sim,
\end{eqnarray} 
where $\sim$ is an equivalence relation from restriction.
\end{itemize}

\subsection{Manifold}
For manifold, in the neighborhood $U_{x_{0}}$ of $x_{0}$, there is coordinate
\begin{eqnarray}
\left.dx^{\mu}\right|_{x\rightarrow x_{0}}.
\end{eqnarray}
As an example of presheaf, the collection of all possible coordinates in the neighborhood $U_{x_{0}}$ of $x_{0}$ is a presheaf
\begin{eqnarray}
d(U_{x_{0}})=\{ \left.dx^{\mu}\right|_{x\rightarrow x_{0}} \}.
\end{eqnarray}
The presheaf $d(U_{x_{0}})$ can gluon to sheaf $d(\mathbf{X})$ because
\begin{eqnarray}
\xymatrix{
d(U_{x_{0}}) \ar[rr]^{\cap d(U_{x_{1}})} & & d(U_{x_{0}}\cap U_{x_{1}})\\
&&\\
U_{x_{0}}\ar[rr]^{\cap U_{x_{1}}} \ar[uu]_{d}& & U_{x_{0}}\cap U_{x_{1}} \ar[uu]_{d}
}
\end{eqnarray}
$d$ is a functor from topological space $\mathbf{X}$ to the differential structure $d(\mathbf{X})$ of the topological space $\mathbf{X}$ 
\begin{eqnarray}
d:\mathbf{X}\rightarrow d(\mathbf{X}).
\end{eqnarray}
{where differential structure $d(\mathbf{X})$ is collection of all possible global coordinates on topological space $\mathbf{X}$,}
$d$ is one kind of functor $F$, and $d(\mathbf{X})$ is one kind of sheaf $F(\mathbf{X})$.
Then we can see the manifold $\mathbf{M}$ is topological space $\mathbf{X}$ with differential structure $d(\mathbf{X})$ ( the global coordinates {on $(1+n)$-dimensional manifold $\mathbf{M}$} might not be parameterized by $\mathbb{R}^{1+n}$)
\begin{eqnarray}
\mathbf{M}=(\mathbf{X}, d(\mathbf{X})).
\end{eqnarray}
\begin{itemize}
\item We {point out} that the definition of manifold $\mathbf{M}=(\mathbf{X}, d(\mathbf{X}))$ is equivalent with the definition in usual book with axioms of locally flatness and atlas compatibility. 
\item (Locally flatness axiom) The point $x_{0}$ in $(1+n)$-dimensional manifold, then the neighborhood $U_{x_{0}}$ can isomorphic to $\mathbb{R}^{1+n}$.
\item (Atlas compatibility axiom) The points $x_{0}$ and $x_{1}$ in $(1+n)$-dimensional manifold have neighborhood $U_{x_{0}}$ and $U_{x_{1}}$ with {parametrization} $\{x_{0}^{\mu}, \mu=0,1,2,\cdots, n\}$ and $\{x_{1}^{\mu}, \mu=0,1,2,\cdots, n\}$. Then, there are coordinates $\{dx_{0}^{\mu}, \mu=0,1,2,\cdots, n$ and $\{dx_{0}^{\mu}, \mu=0,1,2,\cdots, n\}$. For the overlap of the two neighborhood 
\begin{eqnarray}
U_{x_{0}}\cap U_{x_{1}},
\end{eqnarray}
there is coordinate transformation
\begin{eqnarray}
dx^{\mu}_{0}=\Lambda^{\mu}_{\ \nu}(x_{0}) dx^{\nu}_{1}=\Lambda^{\mu}_{\ \nu}(x_{1}) dx^{\nu}_{1}, &\quad& \Lambda^{\mu}_{\ \nu}(x_{0}), \Lambda^{\mu}_{\ \nu}(x_{1})\in GL(1+n,\mathbb{R}),
\end{eqnarray}
where
\begin{eqnarray}
\Lambda^{\mu}_{\ \nu}(x_{0})=\left.\Lambda^{\mu}_{\ \nu}(x)\right|_{x\rightarrow x_{0}}.
\end{eqnarray}

\item For any number of neighborhood, there is 
\begin{eqnarray}\label{differential_structure}
dx^{\mu}_{0}=\prod \left(\Lambda^{\mu}_{\ \nu_{1}}(x_{0})\Lambda^{\nu_{1}}_{\ \nu_{2}}(x_{1})\cdots \Lambda^{\nu_{q-1}}_{\ \nu_{q}}(x_{q}) \right) dx^{\nu_{q}}_{q}\in Hom\left(\Lambda^{\mu}_{\ \nu}(x_{0}),\Lambda^{\mu}_{\ \nu}(x_{q})\right) dx^{\nu}_{q}, \ \ \ 
\end{eqnarray}
where the element in $Hom\left(\Lambda^{\mu}_{\ \nu}(x_{0}),\Lambda^{\mu}_{\ \nu}(x_{q})\right)$ is path dependent and cotangent principal bundle section dependent element of linear transformation group $GL(n,\mathbb{R})$ valued. Then
\begin{eqnarray}\label{coordinate_equation}
x^{\mu}_{0}-C_{q}+C_{0}\in Hom\left(\Lambda^{\mu}_{\ \nu}(x_{0}),\Lambda^{\mu}_{\ \nu}(x_{q})\right) x^{\nu}_{q}.
\end{eqnarray}
Which means the parameters in $x_{0}$ and $x_{q}$ relies on the continues path 
\begin{eqnarray}
C:\tau \rightarrow \mathbf{M}, \quad {C(\tau^{k})=x^{k}}, \quad k=0,1,2,\cdots, n,
\end{eqnarray} 
linear transformation 
\begin{eqnarray}
Hom\left(\Lambda^{\mu}_{\ \nu}(x_{0}),\Lambda^{\mu}_{\ \nu}(x_{q})\right)
\end{eqnarray}
and the edge function $C_{q}-C_{0}$. The solutions of equation (\ref{differential_structure}) just the sheaf space $d(\mathbf{X})$ restricted on curve $C(\tau)$
\begin{eqnarray}
\left.d(\mathbf{X})\right|_{C(\tau)}.
\end{eqnarray}
From equation (\ref{coordinate_equation}) we can see that the {global} coordinates in $(1+n)$-dimensional manifold might not parameterized by $\mathbb{R}^{1+n}$.
\end{itemize}

\subsection{Category viewing of manifold}
\begin{itemize}
\item The manifold $\mathbf{M}$ is a category with objects
\begin{eqnarray}
U_{x_{0}}, d(U_{x_{0}}) \in ob(\mathbf{M}), 
\end{eqnarray}
and morphisms
\begin{eqnarray}
\subset, \cup,\cap, d \in hom(\mathbf{M}). 
\end{eqnarray}
\item The category $\mathbf{Man}$ with objects
\begin{eqnarray}
\mathbf{M}\in ob(\mathbf{Man}),
\end{eqnarray}
and morphisms
\begin{eqnarray}
continuous\ \ differentiable\ \ map \in hom(\mathbf{Man}).
\end{eqnarray}
\end{itemize}
\subsection{Principal bundle}
The fiber $E(U_{x_{0}})$ of the cotangent principal bundle $E(\mathbf{M})$ on manifold $\mathbf{M}$ isomorphic to the freedom $G=GL(1+n,\mathbb{R})$ of coordinates can make transformation (left action) locally
\begin{eqnarray}
E(U_{x_{0}})=\{\left. \Lambda^{\mu}_{\ \nu}(x)\right|_{x\rightarrow x_{0}} | \left.dx^{\mu\prime}\right|_{x^{\prime}\rightarrow x_{0}}=\left.\Lambda^{\mu}_{\ \nu}(x)dx^{\nu}\right|_{x\rightarrow x_{0}}, \left.\Lambda^{\mu}_{\ \nu}(x)\right|_{x\rightarrow x_{0}}\in GL(1+n,\mathbb{R}) \}.\ \ \
\end{eqnarray}
The cotangent principal $G$-bundle $E(\mathbf{M})$ on manifold $\mathbf{M}$ is
\begin{eqnarray}
E(\mathbf{M})=\cup E(U_{x}), &\quad& G=GL(1+n,\mathbb{R}),
\end{eqnarray}
so the cotangent principal bundle is a map $\pi$ from total space $E$ to base manifold $\mathbf{M}$
\begin{eqnarray}
\pi:E\rightarrow \mathbf{M}.
\end{eqnarray}
\begin{eqnarray}
\xymatrix{
E(U_{x_{0}}) \ar[rr]^{\cup E(U_{x})} \ar[dd]^{\pi} & & E(U_{x_{0}}\cup U_{x}) \ar[rr]^{\cup} \ar[dd]^{\pi}&& E(\mathbf{M})\ar[dd]^{\pi} \\
& & & & \\
U_{x_{0}} \ar[rr]^{\cup U_{x}} & & U_{x_{0}}\cup U_{x} \ar[rr]^{\cup}&&\mathbf{M}
}
\end{eqnarray}
The inverse mapping of $\pi$ is a section of the sheaf $d(\mathbf{M})$ and bundle $E(\mathbf{M})$ 
\begin{eqnarray}
\pi^{-1}\in d(\mathbf{M}), \quad \pi^{-1}\subset {E(\mathbf{M})}.
\end{eqnarray}
Here for inverse mapping of cotangent principal bundle, the meaning of $\pi^{-1}$ is one global {coordinate} of the manifold $\mathbf{M}$. 
The contravariant functor $\hat{\pi}^{-1}$ of $\pi$ is the differential structure sheaf of the manifold $\mathbf{M}$ (all possible global coordinates). Because we have the commutative diagram
\begin{eqnarray}
\xymatrix{
d(U_{x_{0}}) \ar[rr]^{\cup d(U_{x})} & & d(U_{x_{0}}\cup U_{x}) \ar[rr]^{\cup} && d(\mathbf{M}) \\
& & & & \\
U_{x_{0}} \ar[rr]^{\cup U_{x}} \ar[uu]^{\hat{\pi}^{-1}}_{d}& & U_{x_{0}}\cup U_{x} \ar[rr]^{\cup} \ar[uu]^{\hat{\pi}^{-1}}_{d} &&\mathbf{M}
\ar[uu]^{\hat{\pi}^{-1}}_{d} }
\end{eqnarray}
then
\begin{eqnarray}
\hat{\pi}^{-1}=d.
\end{eqnarray}

The tangent principal bundle $E^{*}(\mathbf{M})$ is the dual bundle of cotangent principal bundle $E(\mathbf{M})$
\begin{eqnarray}
\pi^{*}:E^{*}\rightarrow \mathbf{M}.
\end{eqnarray}
The section $\pi^{*-1}$ in the neighborhood of $U_{x_{0}}$ has the formula
\begin{eqnarray}
\left.\frac{\partial}{\partial x^{\mu}}\right|_{x\rightarrow x_{0}}
\end{eqnarray} 
and dual with coordinates
\begin{eqnarray}\label{dual}
\left.\langle dx^{\mu} ,\frac{\partial}{\partial x^{\nu}} \rangle\right|_{x\rightarrow x_{0}}=\delta^{\mu}_{\nu}.
\end{eqnarray}
The sheaf $\hat{\pi}^{*-1}$ is dual with $\hat{\pi}^{-1}$. The right action of element of $GL(1+n,\mathbb{R})$ on tangent principal bundle is consistent with the definition of left action transformation on cotangent bundle
\begin{eqnarray}
\left(\frac{\partial}{\partial x^{\mu}}\right)^{\prime}=\frac{\partial}{\partial x^{\nu}}\Lambda_{\ \mu}^{\nu}(x).
\end{eqnarray}
The definition of dual basis (\ref{dual}) gives us that
\begin{eqnarray}
\left.\Lambda^{\mu}_{\ \rho}(x)\Lambda_{\ \nu}^{\rho}(x)\right|_{x\rightarrow x_{0}}=\delta^{\mu}_{\nu}.
\end{eqnarray}

\subsection{Principal bundle connection}
For a section $\pi^{-1}$ of the cotangent principal bundle fiber $E(U_{x_{0}})$ on manifold $\mathbf{M}$, the linear connection operator $\nabla_{\rho}$ is
\begin{eqnarray}
\nabla_{\rho} dx^{\mu}=\left.\frac{dx^{\mu}-\Lambda^{\mu}_{\ \nu}(x_{0})dx^{\nu}_{0}}{x^{\rho}-x^{\rho}_{0}}\right|_{x\rightarrow x_{0}}=\frac{(\delta^{\mu}_{\nu}-\Lambda^{\mu}_{\ \nu}(x))dx^{\nu}}{dx^{\rho}}= -\Gamma^{\mu}_{\ \nu\rho}(x)dx^{\nu},
\end{eqnarray}
then the linear connection operator $\nabla_{\rho}$ is a functor connects fiber $E(U_{x})$ to $E(U_{x_{0}})$
\begin{eqnarray}
\nabla_{\rho}:E(U_{x})\rightarrow E(U_{x_{0}}),\quad x\rightarrow x_{0}.
\end{eqnarray} 
We write connection $1$-form as follow
\begin{eqnarray}
\Gamma^{\mu}_{\ \nu}(x)=\Gamma^{\mu}_{\ \nu\rho}(x)dx^{\rho},
\end{eqnarray}
and the linear connection $1$-form operator
\begin{eqnarray}
\nabla=\nabla_{\rho}dx^{\rho},\quad \nabla dx^{\mu}=-\Gamma^{\mu}_{\ \nu}(x)dx^{\nu}.
\end{eqnarray}
The section of the fiber $E^{*}(U_{x_{0}})$ of tangent bundle has the connection
\begin{eqnarray}
\nabla_{\rho}\left(\frac{\partial}{\partial x^{\mu}}\right)= \frac{\partial}{\partial x^{\nu}}\tilde{\Gamma}_{\ \mu\rho}^{\nu}(x).
\end{eqnarray}
We omit the $x$ index some places below. The dual relation (\ref{dual}) of bases gives us that
\begin{eqnarray}
\nabla_{\rho}\langle dx^{\mu} ,\frac{\partial}{\partial x^{\nu}} \rangle=0,&\Rightarrow &\tilde{\Gamma}_{\ \nu\rho}^{\mu}(x)=-\Gamma_{\ \nu\rho}^{\mu}(x),
\end{eqnarray}
then 
\begin{eqnarray}
\nabla_{\rho}\left(\frac{\partial}{\partial x^{\mu}}\right)= \Gamma_{\ \mu\rho}^{\nu}(x)\frac{\partial}{\partial x^{\nu}}.
\end{eqnarray}
We assume that the linear connection operator $\nabla_{\rho} $ can be defined globally on the manifold $\mathbf{M}$.

Under the coordinate transformation in the neighborhood $U_{x}$, the transformation rule of the principal bundle connection is derived
\begin{eqnarray}\label{connection_one} \nonumber
\nabla_{\rho} dx^{\prime \mu}&=& \nabla_{\rho} \left(\Lambda^{\mu}_{\ \nu}(x) dx^{\nu} \right),\\ \nonumber
\Rightarrow -\Gamma^{\prime\mu}_{\ \nu\rho}(x^{\prime})dx^{\prime\nu}&=&\left(\frac{\partial \Lambda^{\mu}_{\ \sigma}(x)}{\partial x^{\rho}}-\Lambda^{\mu}_{\ \nu}(x)\Gamma^{\nu}_{\ \sigma\rho}(x)\right)dx^{\sigma},\\ \nonumber
\Rightarrow \Gamma^{\prime\mu}_{\ \nu\rho}(x^{\prime})\Lambda^{\nu}_{\ \sigma}(x)&=&\left(\Lambda^{\mu}_{\ \nu}(x)\Gamma^{\nu}_{\ \sigma\rho}(x)-\frac{\partial \Lambda^{\mu}_{\ \sigma}(x)}{\partial x^{\rho}}\right),\\
\Rightarrow \Gamma^{\prime\mu}_{\ \nu\rho}(x^{\prime})&=&\left(\Lambda^{\mu}_{\ \nu}(x)\Gamma^{\nu}_{\ \sigma\rho}(x)-\frac{\partial \Lambda^{\mu}_{\ \sigma}(x)}{\partial x^{\rho}}\right)\Lambda^{\sigma}_{\ \nu}(x),\\
\Rightarrow \Gamma^{\prime\mu}_{\ \nu}(x^{\prime})&=&\left(\Lambda^{\mu}_{\ \nu}(x)\Gamma^{\nu}_{\ \sigma}(x)-d\Lambda^{\mu}_{\ \sigma}(x)\right)\Lambda^{\sigma}_{\ \nu}(x),
\end{eqnarray}
such that the cotangent principal bundle $E(\mathbf{M})$ has structure of connection preserving left action $G=GL(1+n,\mathbb{R})$ torsors
\begin{eqnarray}
E(\mathbf{M})=\frac{G\times E(\mathbf{M})}{G}.
\end{eqnarray}


\subsection{Tangent and cotangent associated bundle}
The tangent associated bundle $TE^{*}(\mathbf{M})$ on manifold $\mathbf{M}$ is glued with tangent space on neighborhood of $x$
\begin{eqnarray}
p^{*}:TE^{*}\rightarrow \mathbf{M},&\quad& TE^{*}(\mathbf{M})=\cup TE^{*}(U_{x}),
\end{eqnarray} 
the section $p^{*-1}$ of the bundle is a vector field of manifold $\mathbf{M}$
\begin{eqnarray}
V(x)=V^{\mu}(x)\frac{\partial}{\partial x^{\mu}},&\quad & \left.(V^{0}(x),V^{1}(x), \cdots, V^{n}(x))\right|_{x\rightarrow x_{0}}\in \mathbb{R}^{1+n}.
\end{eqnarray}
Then the fiber $TE^{*}(U_{x_{0}})$ of the bundle $TE^{*}(\mathbf{M})$ is isomorphic to $\mathbb{R}^{1+n}$. For definite section of the tangent bundle, there is $GL(1+n,\mathbb{R})$ freedom to choose the bases of vector in the neighborhood of $x_{0}$
\begin{eqnarray}\label{free_G}
\left.V(x)\right|_{x\rightarrow x_{0}}=\left.V^{\mu\prime}(x)\frac{\partial }{\partial x^{\nu}}\Lambda^{\nu}_{\ \mu}(x)\right|_{x\rightarrow x_{0}}=\left.V^{\nu}(x)\frac{\partial }{\partial x^{\nu}}\right|_{x\rightarrow x_{0}}, & & \left.\Lambda^{\nu}_{\ \mu}(x)\right|_{x\rightarrow x_{0}} \in GL(1+n,\mathbb{R}).\ \ \ \ \ \ \ \ 
\end{eqnarray}
With the help of (\ref{connection_one}) and (\ref{free_G}), the tangent associated bundle $TE^{*}(\mathbf{M})$ has the structure of connection preserving right action $G=GL(1+n,\mathbb{R})$ torsors
\begin{eqnarray}
TE^{*}(\mathbf{M})=\frac{TE^{*}(\mathbf{M})\times G}{G},
\end{eqnarray}
the right action structure group $G$ of {tangent associated bundle} is free and transitive. The contravariant functor $\hat{p}^{*-1}$ of tangent associated bundle $TE^{*}(\mathbf{M})$ is a sheaf on manifold $\mathbf{M}$
\begin{eqnarray}
\hat{p}^{*-1}:\mathbf{M}\rightarrow TE^{*},
\end{eqnarray}
where the sheaf $\hat{p}^{*-1}$ are collections of all tangent vector fields on manifold $\mathbf{M}$. The sheaf $\hat{p}^{*-1}$ has structure of connection preserving right action $G=GL(1+n,\mathbb{R})$ torsors
\begin{eqnarray}
\hat{p}^{*-1}=\frac{\hat{p}^{*-1}\times G}{G}.
\end{eqnarray} 

Similarly, the cotangent associated bundle is
\begin{eqnarray}
p:TE\rightarrow \mathbf{M},&\quad& TE(\mathbf{M})=\cup TE(U_{x}),
\end{eqnarray}
and has the structure of connection preserving left action $G=GL(1+n,\mathbb{R})$ torsors
\begin{eqnarray}
TE(\mathbf{M})=\frac{ G\times TE(\mathbf{M})}{G},
\end{eqnarray} 
the section $p^{-1}$ of the cotangent associated bundle is cotangent vector field ($1$-form) on manifold $\mathbf{M}$
\begin{eqnarray}
\alpha(x)=\alpha_{\mu}(x) dx^{\mu},&\quad& \left.(\alpha_{0}(x), \alpha_{1}(x),\cdots, \alpha_{n}(x) )\right|_{x\rightarrow x_{0}} \in \mathbb{R}^{1+n}. 
\end{eqnarray}
The contravariant functor $\hat{p}^{-1}$ is the sheaf of all cotangent vector field on manifold $\mathbf{M}$
\begin{eqnarray}
\hat{p}^{-1}:\mathbf{M}\rightarrow TE.
\end{eqnarray}
The sheaf $\hat{p}^{-1}$ have structure of connection preserving left action $G=GL(1+n,\mathbb{R})$ torsors
\begin{eqnarray}
\hat{p}^{-1}=\frac{G\times \hat{p}^{-1}}{G}.
\end{eqnarray}

\section{Lorentz manifold, Riemann geometry and Cartan geometry}

%
%

\subsection{Metric}
Pseudo Riemann geometry 
\begin{eqnarray}
\mathbf{pR}=(\mathbf{M},g)
\end{eqnarray}
is one of most successful geometry system. The pseudo Riemann geometry $\mathbf{pR}$ is a differentiable manifold $\mathbf{M}$ with smooth metric tensor $g$
\begin{eqnarray}
g(x) = -g_{\mu \nu}(x){{dx^{\mu}}\otimes{dx^{\nu}}},
\end{eqnarray}
the metric is symmetric two rank tensor field on manifold $\mathbf{M}$ such that the components of metric tensor
\begin{eqnarray}
g_{\mu \nu}(x)= g_{\nu \mu}(x),
\end{eqnarray}
the metric field is non-degenerate, which means, the determinants of metric tensor components at any point $x_{0}$ in manifold $\mathbf{M}$ are not zero
\begin{eqnarray}
\left. g_{v}\right|_{x\rightarrow x_{0}}=\left.\det (g_{\mu \nu}(x))\right|_{x\rightarrow x_{0}}\neq 0.
\end{eqnarray}

The pseudo Riemann manifold $\mathbf{pR}$ has corresponding inverse metric
\begin{eqnarray}
g^{-1}(x)=-g^{\mu\nu}(x)\frac{\partial}{\partial x^{\mu}}\frac{\partial}{\partial x^{\nu}},
\end{eqnarray}
where the dual basis $\left.\frac{\partial}{\partial x^{\mu}}\right|_{x\rightarrow x_{0}}$ of coordinate $\left.dx^{\mu}\right|_{x\rightarrow x_{0}}$ in the neighborhood of $x_{0}$ satisfy the inner product relation with coordinate 
\begin{eqnarray}\label{dual_coordinate}
\left.\langle \partial_{\mu},dx^{\nu}\rangle\right|_{x\rightarrow x_{0}}=\delta^{\mu}_{\nu}.
\end{eqnarray}
The components of inverse metric $g^{\mu\nu}(x)$ are inverse matrix of metric components $g_{\mu\nu}(x)$ in any point $x_{0}$
\begin{eqnarray}\label{metric_component}
\left.g^{\mu\nu}(x)g_{\nu\rho}(x)\right|_{x\rightarrow x_{0}}=\delta^{\mu}_{\rho}.
\end{eqnarray}
The metric is compatible with linear connection when
\begin{eqnarray}\label{compatible}
\nabla g(x)=0,\\
\Rightarrow \frac{\partial g_{\mu\nu}(x)}{\partial x^{\rho}}-g_{\mu\sigma}(x)\Gamma^{\sigma}_{\ \nu\rho}(x)
-g_{\sigma\nu}(x)\Gamma^{\sigma}_{\ \mu\rho}(x)=0.
\end{eqnarray}
We discuss the {$(1+n)$}-dimensional pseudo Riemann manifold $\mathbf{pR}$ with signature $(-,+,+,\cdots)$, Lorentz manifold $\mathbf{L}$, and with signature $(-,-,-,\cdots)$, Riemann manifold $\mathbf{R}$ 
\begin{eqnarray}
\mathbf{L},\mathbf{R}\subset \mathbf{pR}.
\end{eqnarray} 
Then, $x=(x^{\mu})=(x^{0},x^{q})=(t,\vec{x}), (q=1,2,\cdots,n)$ parameterized the (1+n)-dimensional manifold $\mathbf{L}$ and $\mathbf{R}$, and $dx^{\mu}|_{x\rightarrow x_{0}} (\mu=0,1,2,\cdots,n)$ is a coordinate in the neighborhood of $x_{0}$. 

\subsection{Curve on Lorentz manifold and Riemann manifold}
The curve $C(\tau)$ on manifold $\mathbf{L}$ and $\mathbf{R}$ is defined
\begin{eqnarray}
C:\tau\rightarrow \mathbf{L},\mathbf{R}, &\quad & \tau \in \mathbb{R}.
\end{eqnarray}
The curve $C(\tau)$ on manifold $\mathbf{L}$ and $\mathbf{R}$ is an entity then the curve $C(\tau)$ satisfy the reparameterization symmetry
\begin{eqnarray}
\xymatrix{
\tau \ar[rr]^{f}\ar[dr]_{C} & & f(\tau) \ar[dl]^{C^{\prime}} \\
& \mathbf{L},\mathbf{R} &
}
\end{eqnarray}
then 
\begin{eqnarray}
C(\tau)=C^{\prime}(f(\tau)), &\quad & \tau,f(\tau)\in {\mathbb{R}}.
\end{eqnarray}
The metric $g(x)$ on manifold $\mathbf{L}$ and $\mathbf{R}$ defines a line element of the curve $C(\tau)$
\begin{eqnarray}
ds=\sqrt{-g_{\mu\nu}\frac{d x^{\mu}}{d\tau}\frac{d x^{\nu}}{d\tau}} d\tau.
\end{eqnarray} 
The length of the any path $C(\tau)$ from $x_{0}$ point to $x_{q}$ point on manifold $\mathbf{L}$ and $\mathbf{R}$ is defined
\begin{eqnarray}
s=\int_{x_{0}}^{x_{q}} ds=\int_{x_{0}}^{x_{q}} \sqrt{-g_{\mu\nu}\frac{d x^{\mu}}{d\tau}\frac{d x^{\nu}}{d\tau}} d\tau.
\end{eqnarray}
The variation of the length $s$ from point $x_{0}$ to $x_{q}$ screen out the geodesic curve from point $x_{0}$ to $x_{q}$ on manifold $\mathbf{L}$ and $\mathbf{R}$
\begin{eqnarray}\label{geodesic}
\delta s=0.
\end{eqnarray}
The definition (\ref{geodesic}) of geodesic curve derives that
\begin{eqnarray}
\frac{d^{2} x^{\mu} }{d\tau^{2}}+\left\{ ^{\mu}_{\ \nu\rho} \right\}\frac{dx^{\nu}}{d\tau} \frac{dx^{\rho}}{d\tau}=0,
\end{eqnarray} 
the $\left\{ ^{\mu}_{\ \nu\rho}\right\} $ is Christoffel symbol and defined by metric components
\begin{eqnarray}
\left\{ ^{\mu}_{\ \nu\rho} \right\}=\frac{1}{2}g^{\mu\sigma}(\frac{\partial g_{\sigma\rho}}{\partial x^{\nu}}+\frac{\partial g_{\sigma\nu}}{\partial x^{\rho}}-\frac{\partial g_{\mu\nu}}{\partial x^{\sigma}}).
\end{eqnarray}
The $d\tau$ is basis of cotangent vector on curve $C(\tau)$, and the dual basis $\frac{d}{d\tau}$ is defined 
\begin{eqnarray}
\left.\langle d\tau ,\frac{d}{d\tau} \rangle\right|_{\tau=\tau_{0}}=1.
\end{eqnarray} 
The restriction of tangent principal bundle $E^{*}$ from manifold $\mathbf{L}$ and $\mathbf{R}$ to curve $C(\tau)$ is
\begin{eqnarray}
\xymatrix{
E^{*}(\mathbf{L}),E^{*}(\mathbf{R}) \ar[rr]^{restriction} \ar[dd]^{\pi} & & E^{*}(U_{\tau}) \ar[dd]^{\pi} \\
& & \\
\mathbf{L},\mathbf{R} \ar[rr]^{restriction} & & U_{\tau} 
}
\end{eqnarray}
The objects in $E^{*}(U_{\tau})$ are tangent vector on the curve $C(\tau)$
\begin{eqnarray}
\frac{d}{d\tau}\in E^{*}(U_{\tau}),&\quad & \tau\in \mathbb{R}.
\end{eqnarray}
When the linear connection operator $\nabla_{\rho}$ acting on tangent vector $\frac{d}{d\tau}$ of curve $C(\tau)$ equals zero, the curve $C(\tau)$ is self-parallel transported
\begin{eqnarray}\label{self_parallel}
\nabla_{\rho}\left(\frac{d}{d\tau}\right)=0.
\end{eqnarray} 
The definition of self-parallel (\ref{self_parallel}) derives that
\begin{eqnarray}
\frac{d^{2} x^{\mu} }{d\tau^{2}}+\Gamma ^{\mu}_{\ \nu\rho}(\tau) \frac{dx^{\nu}}{d\tau} \frac{dx^{\rho}}{d\tau}=0.
\end{eqnarray}

\subsection{Principal bundle on Lorentz manifold and Riemann manifold}
The freedom to choose $dx^{\mu}|_{x\rightarrow x_{0}}$ is isomorphic to the fiber $E(U_{x_{0}})$ of the cotangent principal bundle $E(\mathbf{L})$ and $E(\mathbf{R})$ of the (1+n)-dimensional Lorentz manifold $\mathbf{L}$ and Riemann manifold $\mathbf{R}$. There is freedom to choose coordinate in the neighborhood of $x_{0}$
\begin{eqnarray}
E(U_{x_{0}})=\left.\{\left.\Lambda^{\mu}_{\ \nu}(x)\right|_{x\rightarrow x_{0}} \right| \left.dx^{\mu\prime}\right|_{x^{\prime}\rightarrow x_{0}}=\left.\Lambda^{\mu}_{\ \nu}(x)dx^{\nu}\right|_{x\rightarrow x_{0}}, \left.\Lambda^{\mu}_{\ \nu}(x)\right|_{x\rightarrow x_{0}}\in GL(1+n,\mathbb{R}) \},\ \ \ 
\end{eqnarray}
such that the cotangent principal bundle 
\begin{eqnarray}
E(\mathbf{L})=\cup E(U_{x}), \quad x\in \mathbf{L},
\end{eqnarray}
and 
\begin{eqnarray}
E(\mathbf{R})=\cup E(U_{x}), \quad x\in \mathbf{R},
\end{eqnarray}
has the structure of connection preserving left action $G=GL(1+n,\mathbb{R})$ torsors
\begin{eqnarray}
E(\mathbf{L})=\frac{G\times E(\mathbf{L})}{G},\quad E(\mathbf{R})=\frac{G\times E(\mathbf{R})}{G}.
\end{eqnarray}

For definite metric $g(x)$ of manifold $\mathbf{L}$ and $\mathbf{R}$, there is $GL(1+n,\mathbb{R})$ freedom to choose the coordinate $\left.dx^{\mu}\right|_{x\rightarrow x_{0}}$ locally to describe the same metric $g(x)$ in the neighborhood of $x_{0}$
\begin{eqnarray}\nonumber
\left.g(x)\right|_{x\rightarrow x_{0}}=-\left.g^{\prime}_{\mu\nu}(x)dx^{\mu\prime} dx^{\nu\prime}\right|_{x\rightarrow x_{0}}=-\left.g^{\prime}_{\mu\nu}(x)\Lambda^{\mu}_{\ \rho}(x)\Lambda^{\nu}_{\ \sigma}(x) dx^{\rho} dx^{\sigma}\right|_{x\rightarrow x_{0}}\\
=-\left.g_{\rho\sigma}(x)dx^{\rho} dx^{\sigma}\right|_{x\rightarrow x_{0}}.
\end{eqnarray}
where 
\begin{eqnarray}\nonumber
\left. g^{\prime}_{\mu\nu}(x)\Lambda^{\mu}_{\ \rho}(x)\Lambda^{\nu}_{\ \sigma}(x) \right|_{x\rightarrow x_{0}}=\left.g_{\rho\sigma}(x)\right|_{x\rightarrow x_{0}}
\end{eqnarray}
are used.

For inverse metric, the analyze for tangent principal bundles $E^{*}(\mathbf{L})$ and $E^{*}(\mathbf{R})$ are similar, and the tangent principal bundle on (1+n)-dimensional manifold $\mathbf{L}$ and $\mathbf{R}$ has structure of connection preserving right $G=GL(1+n,\mathbb{R})$ action torsors
\begin{eqnarray}
E^{*}(\mathbf{L})=\frac{E^{*}(\mathbf{L})\times G}{G},\quad E^{*}(\mathbf{R})=\frac{E^{*}(\mathbf{R})\times G}{G}.
\end{eqnarray}

\subsection{Orthogonal principal frame bundle and Cartan geometry}
The inverse metric $g^{-1}(x)$ in Lorentz manifold $\mathbf{L}$ is described by orthonormal frame formalism ($a,b=0,1,2,\cdots,n$)
\begin{equation}\label{g}
g^{-1}(x) =-{\eta }^{ab}{{\theta }_{a}}(x){{\theta }_{b}}(x), 
\end{equation}
where 
\begin{eqnarray}
\eta^{ab}=diag(1,-1,-1,\cdots,-1)
\end{eqnarray}
and
\begin{eqnarray}
{\theta }_{a}(x)={\theta }_{a}^{\mu}(x) \frac{\partial}{\partial x^{\mu}}
\end{eqnarray}
are orthonormal frames and describe gravitational field. The Riemann manifold $\mathbf{R}$ be described by inverse metric $\bar{g}^{-1}(x)$ orthonormal frame formalism as 
\begin{equation}\label{barg}
\bar{g}^{-1}(x) =-{I}^{ab}{{\theta }_{a}}(x){{\theta }_{b}}(x), 
\end{equation}
where 
\begin{eqnarray}
I^{ab}=diag(1,1,1,\cdots,1).
\end{eqnarray}
For definite inverse metric $g^{-1}(x)$, there is $O(1,n)$ freedom to choose the orthonormal frame $\left.\theta^{a}(x)\right|_{x\rightarrow x_{0}}$ to describe the same metric in the neighborhood of $x_{0}$
\begin{eqnarray}
\left.\theta^{a\prime}(x)\right|_{x\rightarrow x_{0}}=\left.\Lambda^{a}_{\ b}(x)\theta^{b}(x)\right|_{x\rightarrow x_{0}},&\quad & \left.\Lambda^{a}_{\ b}(x)\right|_{x\rightarrow x_{0}}\in O(1,n),
\end{eqnarray}
and
\begin{eqnarray}\nonumber
\left.g^{-1}(x)\right|_{x\rightarrow x_{0}}=-\left.\eta^{\prime}_{ab}\theta^{a\prime}(x)\theta^{b\prime}(x)\right|_{x\rightarrow x_{0}}=-\left.\eta^{\prime}_{ab}\theta^{c}(x)\Lambda^{a}_{\ c}(x)\theta^{d}(x)\Lambda^{b}_{\ d}(x)\right|_{x\rightarrow x_{0}}\\
=-\left.\eta_{ab}\theta^{a}(x)\theta^{b}(x)\right|_{x\rightarrow x_{0}}, 
\end{eqnarray}
where
\begin{eqnarray}
\left.\eta^{\prime}_{ab}\Lambda^{a}_{\ c}(x)\Lambda^{b}_{\ d}(x)\right|_{x\rightarrow x_{0}}
=\left.\eta_{cd}\right|_{x\rightarrow x_{0}},&\quad & \left.\Lambda^{a}_{\ b}(x)\right|_{x\rightarrow x_{0}}\in O(1,n).
\end{eqnarray}
Which means, the fiber $OE^{*}(U_{x_{0}})$ of orthonormal principal frame bundle $OE^{*}(\mathbf{L})$ is isomorphic to the orthonormal frame transformation freedom $G=O(1,n)$ (right action) locally
\begin{eqnarray}
OE^{*}(U_{x_{0}})=\left\{ \left.\Lambda^{b}_{\ a}(x)\right|_{x\rightarrow x_{0}} | \left.\theta_{a}^{\prime}(x^{\prime})\right|_{x^{\prime}\rightarrow x_{0}}=\left.\theta_{b}(x)\Lambda^{b}_{\ a}(x)\right|_{x\rightarrow x_{0}}, \left.\Lambda^{b}_{\ a}(x)\right|_{x\rightarrow x_{0}}\in O(1,n) \right\}.\ \ \ 
\end{eqnarray}
The orthogonal frame principal bundle is
\begin{eqnarray}
OE^{*}(\mathbf{L})=\cup OE^{*}(U_{x}),\quad x \in \mathbf{L},
\end{eqnarray} 
The fiber $OE^{*}(U_{x_{0}})$ of orthonormal principal frame bundle $OE^{*}(\mathbf{R})$ of Riemann manifold $\mathbf{R}$ is isomorphic to the orthonormal frame transformation freedom $G=O(1+n)$ (right action) locally
\begin{eqnarray}
OE^{*}(U_{x_{0}})=\left\{ \left.\Lambda^{b}_{\ a}(x)\right|_{x\rightarrow x_{0}} | \left.\theta_{a}^{\prime}(x^{\prime})\right|_{x^{\prime}\rightarrow x_{0}}=\left.\theta_{b}(x)\Lambda^{b}_{\ a}(x)\right|_{x\rightarrow x_{0}}, \left.\Lambda^{b}_{\ a}(x)\right|_{x\rightarrow x_{0}}\in O(1+n) \right\}.\ \ \ 
\end{eqnarray}
The orthogonal frame principal bundle is
\begin{eqnarray}
OE^{*}(\mathbf{R})=\cup OE^{*}(U_{x}),\quad x \in \mathbf{R}.
\end{eqnarray}

The metric $g(x)$ and $\bar{g}(x)$ can be described by cotangent orthogonal frame (orthogonal co-frame) formalism as follow
\begin{eqnarray}
g(x)=-\eta_{ab}\theta^{a}(x)\theta^{b}(x),\quad \bar{g}(x)=-I_{ab}\theta^{a}(x)\theta^{b}(x),
\end{eqnarray} 
where
\begin{eqnarray}
\eta_{ab}=diag(1,-1,-1,\cdots,-1), \quad I_{ab}=diag(1,1,1,\cdots,1)
\end{eqnarray}
and
\begin{eqnarray}\label{co_frame}
\theta^{a}(x)=\theta^{a}_{\mu}(x)dx^{\mu}
\end{eqnarray}
are cotangent orthogonal frame. It is derived from (\ref{dual_coordinate}) and (\ref{metric_component}) that the cotangent orthogonal frame is dual with tangent orthogonal frame
\begin{eqnarray}\label{inner_basis}
\left.\langle \theta^{a}(x), \theta_{b}(x)\rangle\right|_{x\rightarrow x_{0}}=\delta^{a}_{b}
\end{eqnarray}
and 
\begin{eqnarray}
\left.\theta^{a}_{\mu}(x)\theta_{b}^{\mu}(x)\right|_{x\rightarrow x_{0}}=\delta^{a}_{b},&\quad& \left.\theta^{a}_{\mu}(x)\theta_{a}^{\nu}(x)\right|_{x\rightarrow x_{0}}=\delta_{\mu}^{\nu}.
\end{eqnarray}
From equation (\ref{inner_basis}) we have
\begin{eqnarray}
\left.\Lambda^{a}_{\ c}(x)\Lambda_{\ b}^{c}(x)\right|_{x\rightarrow x_{0}}=\delta^{a}_{b}.
\end{eqnarray}
The structure group of orthogonal co-frame bundles $OE(\mathbf{L})$ and $OE(\mathbf{R})$ are $O(1,n)$ and $O(1+n)$, also.

The orthogonal frame connection coefficients is defined
\begin{eqnarray}
\nabla_{\rho}\theta_{a}(x)=\nabla_{\rho}\left(\theta_{a}^{\mu}(x)\frac{\partial }{\partial x^{\mu}}\right),\\ \nonumber
\Rightarrow \Gamma^{b}_{\ a\rho}(x)\theta_{b}^{\mu}(x)=\frac{\partial \theta_{a}^{\mu}(x)}{\partial x^{\rho}}+\theta_{a}^{\sigma}(x)\Gamma^{\mu}_{ \ \sigma\rho}(x),\\ \label{Catan_pre}
\Rightarrow \Gamma^{b}_{\ a}(x)\theta_{b}^{\mu}(x)=d\theta_{a}^{\mu}(x)+\theta_{a}^{\sigma}(x)\Gamma^{\mu}_{ \ \sigma}(x),\\\label{frame_connection}
\Rightarrow \Gamma^{b}_{\ a\rho}(x)=\left[\frac{\partial \theta_{a}^{\mu}(x)}{\partial x^{\rho}}+\theta_{a}^{\sigma}(x)\Gamma^{\mu}_{ \ \sigma\rho}(x)\right]\theta^{b}_{\mu}(x).
\end{eqnarray}
Eliminating the edge term 
\begin{eqnarray}
\frac{\partial (\theta_{a}^{\mu}(x) \theta^{b}_{\mu}(x))}{\partial x^{\rho}}=\theta_{a}^{\mu}(x) \frac{\partial \theta^{b}_{\mu}(x)}{\partial x^{\rho}}+\frac{\partial \theta_{a}^{\mu}(x) }{\partial x^{\rho}}\theta^{b}_{\mu}(x),
\end{eqnarray}
(\ref{frame_connection}) can be written as
\begin{eqnarray}
\Gamma^{b}_{\ a\rho}(x)= \left[\theta^{b}_{\sigma}(x)\Gamma^{\sigma}_{ \ \mu\rho}(x)-\frac{\partial \theta^{b}_{\mu}(x)}{\partial x^{\rho}}\right]\theta_{a}^{\mu}(x),\\
\Rightarrow \Gamma^{b}_{\ a}(x)= \left[\theta^{b}_{\sigma}(x)\Gamma^{\sigma}_{ \ \mu}(x)-d\theta^{b}_{\mu}(x)\right]\theta_{a}^{\mu}(x).
\end{eqnarray}
The compatible connection condition (\ref{compatible}) for orthogonal frame connection coefficients is
\begin{eqnarray}
\eta^{ac}\Gamma^{b}_{\ c\rho}(x)+\eta^{cb}\Gamma^{a}_{\ c\rho}(x)=0,\\
\Rightarrow \Gamma^{ba}_{\ \ \rho}(x)=-\Gamma^{ab}_{\ \ \rho}(x).
\end{eqnarray}
The connection $1$-form on orthogonal frame is defined
\begin{eqnarray}
\Gamma^{a}_{\ c}(x)=\Gamma^{a}_{\ c\rho}(x)dx^{\rho}, 
\end{eqnarray}
then we have
\begin{eqnarray}
\nabla \theta_{a}(x)&=&\Gamma^{b}_{\ a}(x)\theta_{b}(x),\\
\nabla \theta^{a}(x)&=&-\Gamma^{a}_{\ b}(x)\theta^{b}(x),\\
\Gamma^{ab}(x)&=&-\Gamma^{ba}(x).
\end{eqnarray}

The structure of connection preserving right action $G=O(1,n)$ and $\bar{G}=O(1+n)$ torsors of orthogonal frame principal bundles
\begin{eqnarray}
OE^{*}(\mathbf{L})=\frac{OE^{*}(\mathbf{L})\times G}{G}, \ and \ OE^{*}(\mathbf{R})=\frac{OE^{*}(\mathbf{R})\times \bar{G}}{\bar{G}},
\end{eqnarray}
derives that 
\begin{eqnarray}
\Gamma^{\prime b}_{\ a\rho}(x^{\prime})=\left(\Lambda^{b}_{\ c}(x)\Gamma^{c}_{\ d\rho}(x)-\frac{\partial \Lambda^{b}_{\ d}(x)}{\partial x^{\rho}}\right)\Lambda^{d}_{\ a}(x),\\
\Rightarrow \Gamma^{\prime b}_{\ a}(x^{\prime})=\left(\Lambda^{b}_{\ c}(x)\Gamma^{c}_{\ d}(x)-d\Lambda^{b}_{\ d}(x)\right)\Lambda^{d}_{\ a}(x),\\ \nonumber
\Rightarrow \Gamma^{\prime b}_{\ a}(x^{\prime})\Lambda^{a}_{\ d}(x)=\left(\Lambda^{b}_{\ c}(x)\Gamma^{c}_{\ d}(x)-d\Lambda^{b}_{\ d}(x)\right),\\ \label{curvature_transformation}
\Rightarrow d\Gamma^{\prime b}_{\ a}(x^{\prime})\Lambda^{a}_{\ d}(x)-\Gamma^{\prime b}_{\ a}(x^{\prime})\wedge d\Lambda^{a}_{\ d}(x)=d\Lambda^{b}_{\ c}(x)\wedge \Gamma^{c}_{\ d}(x)+\Lambda^{b}_{\ c}(x)\wedge d\Gamma^{c}_{\ d}(x),
\end{eqnarray}
the curvature $2$-form is defined
\begin{eqnarray}
\Omega^{a}_{\ b}(x)=d\Gamma^{a}_{\ b}(x)+\Gamma^{a}_{\ c}(x)\wedge \Gamma^{c}_{\ b}(x),
\end{eqnarray}
and equation (\ref{curvature_transformation}) derives that the curvature $2$-form satisfy the tensor transformation rule
\begin{eqnarray}
\Omega^{\prime a}_{\ b}(x^{\prime})\Lambda^{b}_{\ d}(x)=\Lambda^{b}_{\ c}(x)\Omega^{c}_{\ d}(x).
\end{eqnarray}
The relation between curvature $2$-form $\Omega^{a}_{\ b}(x)$ and curvature tensor $R^{a}_{\ b\mu\nu}(x)$ is
\begin{eqnarray}
\Omega^{a}_{\ b}(x)=\frac{1}{2}R^{a}_{\ b\mu\nu}(x)dx^{\mu}\wedge dx^{\nu},
\end{eqnarray}
where 
\begin{eqnarray}\label{cur1}
R^{a}_{\ b\mu\nu}(x)=\frac{\partial \Gamma^{a}_{\ b\nu}(x)}{\partial x^{\mu}}-\frac{\partial \Gamma^{a}_{\ b\mu}(x)}{\partial x^{\nu}}+\Gamma^{a}_{\ c\mu}(x)\Gamma^{c}_{\ b\nu}(x)-\Gamma^{a}_{\ c\nu}(x)\Gamma^{c}_{\ b\mu}(x).
\end{eqnarray}

Equation (\ref{co_frame}) bring us that
\begin{eqnarray}\label{dx_mu}
dx^{\mu}=\theta_{a}^{\mu}(x)\theta^{a}(x),
\end{eqnarray}
after the exterior derivative $d$ being acted on equation (\ref{dx_mu}), the Cartan structure equation is derived 
\begin{eqnarray}\nonumber
0=d\left(\theta_{a}^{\mu}(x)\right)\wedge \theta^{a}(x)+\theta_{a}^{\mu}(x)d\left(\theta^{a}(x)\right),\\
\Rightarrow d\theta^{c}(x)=-\Gamma^{c}_{\ b}(x)\wedge \theta^{b}(x)+\Gamma^{c}_{\ \mu\nu}(x)dx^{\nu}\wedge dx^{\mu}.
\end{eqnarray}
The trosion $2$-form is defined
\begin{eqnarray}
T^{c}(x)=\frac{1}{2}T^{c}_{\ \mu\nu}(x)dx^{\mu}\wedge dx^{\nu}=-\Gamma^{c}_{\ \mu\nu}(x)dx^{\nu}\wedge dx^{\mu},
\end{eqnarray} 
then the components of torsion is
\begin{eqnarray}
T^{c}_{\ \mu\nu}(x)=2\Gamma^{c}_{\ [\mu\nu]}(x)=\Gamma^{c}_{\ \mu\nu}(x)-\Gamma^{c}_{\ \nu\mu}(x),
\end{eqnarray} 
and the Cartan structure equation is rewritten as
\begin{eqnarray}\label{Cartan_first}
d\theta^{c}(x)+\Gamma^{c}_{\ b}(x)\wedge \theta^{b}(x)+T^{c}(x)=0.
\end{eqnarray}
It is easy to prove the torsion satisfy the tensor transformation rule.
The exterior derivative $d$ acting on equation (\ref{Cartan_first}) gives us Ricci identity
\begin{eqnarray}\label{Cartan_second_pre}
d\Gamma^{c}_{\ b}(x)\wedge \theta^{b}(x)-\Gamma^{c}_{\ b}(x)\wedge d\theta^{b}(x)+dT^{c}(x)=0,\\ \label{Ricci_identity}
\Rightarrow \Omega^{c}_{\ b}(x)\wedge \theta^{b}(x)+\Gamma^{c}_{\ b}(x)\wedge T^{b}(x)+dT^{c}(x)=0.
\end{eqnarray}
The equation (\ref{Ricci_identity}) is Ricci identity in Cartan geometry with torsion, and the components formulation is
\begin{eqnarray}
R^{a}_{\ [\rho\mu\nu]}(x) +\Gamma^{a}_{\ \sigma[\rho}(x)T^{\sigma}_{\ \mu\nu]}(x)+\partial_{[\rho}T^{a}_{\ \mu\nu]}(x)=0,
\end{eqnarray} 
where
\begin{eqnarray}
\partial_{\rho}=\frac{\partial}{\partial x^{\rho}}.
\end{eqnarray}
The exterior derivative $d$ acting on {Ricci} identity (\ref{Ricci_identity}) derives that the Bianchi identity
\begin{eqnarray}\label{second_bianchi}
d \Omega^{c}_{ d}(x)-\Omega^{c}_{\ b}(x)\wedge \Gamma^{b}_{\ d}(x)+\Gamma^{c}_{\ d}(x)\wedge\Omega^{d}_{\ b}(x)=0,
\end{eqnarray}
the components formulation of the Bianchi identity is
\begin{eqnarray}
\partial_{[\mu} R^{c}_{\ |d|\nu\rho]}(x)-R^{c}_{\ b[\mu\nu}(x)\Gamma^{b}_{\ |d|\rho]}(x)+\Gamma^{c}_{\ d[\mu}(x)R^{d}_{\ |b|\nu\rho]}(x)=0.
\end{eqnarray}

The determinants of metric components in the neighborhood of $x_{0}$ gives us the coordinate free volume element $\theta_{v}(x)$
\begin{eqnarray}\nonumber
\left. g_{v}(x)\right|_{x \rightarrow x_{0}}=\left. det\left[g_{\mu\nu}(x)\right] \right|_{x \rightarrow x_{0}}=det\left[\eta_{ab}\theta^{a}_{\mu}(x)\theta^{b}_{\mu}(x)\right]=-\left.det^{2}\left[\theta^{a}_{\mu}(x)\right]\right|_{x \rightarrow x_{0}}, \\
\Rightarrow \left. \theta_{v}(x)\right|_{x \rightarrow x_{0}}=\left.det\left[\theta^{a}_{\mu}(x)\right]\right|_{x \rightarrow x_{0}}=\left.\sqrt{-g_{v}(x)}\right|_{x \rightarrow x_{0}}
\end{eqnarray}
\subsection{Category viewing of principal bundle on Lorentz manifold and Riemann manifold}
The cotangent principal bundle $E^{*}(\mathbf{L})$ and $E^{*}(\mathbf{R})$ are dual to tangent bundle $E(\mathbf{L})$ and $E(\mathbf{R})$, the orthogonal functor $O$ acting on associatd bundle gives us orthogonal frame bundle and co-frame bundle
\begin{eqnarray}
O:E\rightarrow OE,&\quad &O:E^{*}\rightarrow OE^{*},
\end{eqnarray} 
and the commutative diagram of these four kinds of associated bundle on Lorentz manifold $\mathbf{L}$ and Riemann manifold $\mathbf{R}$ is as follow.
\begin{eqnarray}
\xymatrix{
OE\ar[rr]^{dual}\ar[dd]&&OE^{*}\ar[dd] |\hole \\
&E\ar[rr]^{dual}\ar[lu]_{O}\ar[dd]&&E^{*}\ar[lu]_{O}\ar[dd]^{\pi}\\
\mathbf{L},\mathbf{R}\ar[rr] |\hole &&\mathbf{L},\mathbf{R} \\
&\mathbf{L},\mathbf{R}\ar[rr]\ar[lu]&&\mathbf{L},\mathbf{R}\ar[lu]_{1}\\
}
\end{eqnarray}

\section{Clifford algebra and Dirac matrices}
\subsection{$Cl_{1,n}$ Clifford algebra and Dirac matrices}
The $Cl_{1,n}(\mathbb{R})$ Clifford algebra has $1+n$ generators $\gamma^{a}(a=0,1,2,\cdots,n)$. The Clifford algebra is spanned by the bases as follows 
\begin{equation}
Cl_{1,n}(\mathbb{R})=\textrm{span} \left\{\begin{split}
&C_{1+n}^{0} \ 0-\textrm{vector}: I,\\
&C_{1+n}^{1} \ 1-\textrm{vector}:\gamma^{a_{1}},\\
&C_{1+n}^{2} \ 2-\textrm{vector}:\gamma^{a_{1}}\gamma^{a_{2}},\ \ \ \ \ \ \ \ \ \ \ \ \ \ \ \ \ ({a}_{1}< {a}_{2} < {a}_{3}< \cdots<{a}_{1+n}).\\
&C_{1+n}^{3} \ 3-\textrm{vector}:\gamma^{a_{1}}\gamma^{a_{2}}\gamma^{a_{3}},\\
&\ \ \ \ \ \ \ \ \ \ \ \ \vdots\\
&C_{1+n}^{1+n} \ (1+n)-\textrm{vector}:\gamma^{a_{1}}\gamma^{a_{2}}\gamma^{a_{3}}\cdots \gamma^{a_{1+n}},\\
\end{split} \right.
\end{equation}
The Clifford algebra $Cl_{1,n}(\mathbb{R})$ is $2^{1+n}$-dimensional linear space and
\begin{eqnarray}
Cl_{1,n}(\mathbb{R})=\{ \alpha I+\alpha_{a_{1}}\gamma^{a_{1}}+\alpha_{a_{1}a_{2}}\gamma^{a_{1}}\gamma^{a_{2}}+\cdots+\alpha_{a_{1}a_{2}\cdots a_{1+n}}\gamma^{a_{1}}\gamma^{a_{2}}\cdots \gamma^{a_{1+n}}\},
\end{eqnarray}
where the coefficients before the bases are real valued
\begin{eqnarray}
\alpha, \alpha_{a_{1}}, \alpha_{a_{1}a_{2}},\cdots, \alpha_{a_{1}a_{2}\cdots a_{1+n}} \in \mathbb{R}.
\end{eqnarray}
The matrix representation of generators of Clifford algebra satisfy the restriction
\begin{equation}\label{clifford}
{{\gamma }^{a}}{{\gamma }^{b}}+{{\gamma }^{b}}{{\gamma }^{a}}=2\eta^{ab}I_{k},
\end{equation}
where $I_{k}$ is $k\times k$ identity matrix.
In physics, the Hermiticity conditions for generators of Clifford algebra can be chosen always
\begin{equation}\label{cliffordA}
{{\gamma }^{a}}{{\gamma }^{b\dagger}}+{{\gamma }^{b\dagger}}{{\gamma }^{a}}=2I^{ab}I_{k},
\end{equation}
where $I^{ab}$ is $(1+n)\times (1+n)$ identity matrix.
The minimal faithful matrix representation for $Cl_{1,n}(\mathbb{R})$ gives us the relation
\begin{eqnarray}
2^{1+n}=k\times k\Rightarrow k=\sqrt{2^{1+n}},
\end{eqnarray}
which means, for any matrix representation of generators of Clifford algebra, there is freedom of $U(k)$ to rotate the matrix representation
\begin{eqnarray} 
\gamma^{a\prime}=\psi^{\dagger}\gamma^{a}\psi, \quad \psi\in U(k),
\end{eqnarray}
such that the $ \gamma^{a\prime}$ still the generators of Clifford algebra $Cl_{1,n}(\mathbb{R})$.
The Dirac matrices can be represented by components formula
\begin{eqnarray}
\gamma^{a}=\gamma^{a}_{\ ij}e^{\dagger}_{j}\otimes e_{i}=\psi^{\dagger}_{i}\gamma^{a}\psi_{j}e^{\dagger}_{j}\otimes e_{i}, \quad \psi\in U(k),
\end{eqnarray}
where $e_{i}(i=1,2,\cdots,k)$ are the orthogonal bases expanding (1+n)-dimension complex space $\mathbb{C}^{k}$ and 
\begin{equation}\label{ei}
\mathbf{tr}( e^{\dagger}_{j} \otimes e_{i})=e_{i}e^{\dagger}_{j}=\delta_{ij}.
\end{equation}
One simple choice of $e_{i}$ is
\begin{eqnarray} \label{bases} 
e_{1}=(e^{i\theta_{1}},0,0,\cdots,0), &\quad& e_{2}=(0,e^{i\theta_{2}},0,\cdots,0),\\
\cdots\ \ \ \ \ \ \ \ \ \ \ \ , &\quad& e_{k}=(0,0,0,\cdots,e^{i\theta_{k}}).
\end{eqnarray}
\subsection{$Cl_{1,3}(\mathbb{R})$ Clifford algebra and Dirac matrices}
Particularly, the solution with 
\begin{eqnarray}
n=3 \ and \ 1, \quad k=4\ and \ 2,
\end{eqnarray}
are particular important from the reasons of physics.
The corresponding Clifford algebra are $Cl_{1,3}(\mathbb{R})$ and $Cl_{1,1}(\mathbb{R})$. The generators of Clifford algebra $Cl_{1,3}(\mathbb{R})$ is well know Dirac matrices and the bases
\begin{equation}
Cl_{1,3}(\mathbb{R})=\textrm{span} \left\{\begin{split}
&1 \ \textrm{scalar}: I,\\
&4 \ \textrm{vector}:\gamma^{a},\\
&6 \ \textrm{bivector}:\gamma^{a}\gamma^{b},\ \ \ \ \ \ \ \ \ \ \ \ \ \ \ \ \ ({a}< {b} < {c}< {d}).\\
&4 \ \textrm{pseudovectors}:\gamma^{a}\gamma^{b}\gamma^{c},\\
&1 \ \textmd{pseudoscalar}:\gamma^{a}\gamma^{b}\gamma^{c}\gamma^{d},\\
\end{split} \right.
\end{equation}
The Weyl representation $(q=1,2,3)$ of Dirac matrices are
\begin{eqnarray}
\gamma^{0}=\left(\begin{array}{cc}
0&I_{2\times 2},\\
I_{2\times 2}&0 
\end{array}\right), &\quad & \gamma^{q}=\left(\begin{array}{cc}
0&\sigma_{q},\\
-\sigma_{q}&0 
\end{array}\right),
\end{eqnarray}
where $\sigma_{q}$ are Pauli matrices
\begin{eqnarray}
\sigma_{1}=\left(\begin{array}{cc} 0&1\\
1&0
\end{array}\right),\quad \sigma_{2}=\left(\begin{array}{cc} 0&-i\\
i&0
\end{array}\right), \quad \sigma_{3}=\left(\begin{array}{cc} 1&0\\
0&-1
\end{array}\right).
\end{eqnarray}
The components formulation of $\gamma^{a \prime }$ is
\begin{eqnarray}
\gamma^{a\prime}=\psi^{\dagger}_{li}\gamma^{a}_{\ ij}\psi_{jk}e_{k}\otimes e^{\dagger}_{l}=\psi^{\dagger}_{i}\gamma^{a}_{\ }\psi_{j} e_{j}\otimes e^{\dagger}_{i}, \quad \psi\in U(4),
\end{eqnarray}
with $i,j=1,2,3,4$, where $\psi_{1},\psi_{2},\psi_{3}$ and $\psi_{4}$ are four kinds of Dirac spinors.
The element of $U(4)$ group can be presented
\begin{eqnarray}
\psi=e^{-iV_{\alpha} \mathcal{T}^{\alpha}},\quad V_{\alpha}\in \mathbb{R}, \quad \alpha=0,1,2,\cdots,15.
\end{eqnarray}
The $\mathcal{T}^{\alpha}$ is generators of $U(4)$ group and
\begin{eqnarray}
\mathcal{T}^{\alpha}=\mathcal{T}^{\alpha\dagger}.
\end{eqnarray}
\subsection{$Cl_{1,1}(\mathbb{R})$ Clifford algebra}
The generators of Clifford algebra $Cl_{1,1}(\mathbb{R})$ can be represented by Pauli matrices, as an example
\begin{equation}
Cl_{1,1}(\mathbb{R})=\textrm{span} \left\{\begin{split}
&1 \ \textrm{scalar}: I,\\
&2 \ \textrm{vector}: \gamma^{0}=\sigma_{1}, \quad \gamma^{1}=i\sigma_{2},\\
&1 \ \textrm{bivector}: -\sigma_{3},\\
\end{split} \right.
\end{equation}

\subsection{Isomorphism between bases of $Cl_{1,3}(\mathbb{R})$ and the generators of $U(4)$ group}
An isomorphism between the bases of $Cl_{1,3}(\mathbb{R})$ and the generators of $U(4)$ group can be constructed as follow. The modified Dirac matrices could be
\begin{align*}
\m{T}^{1,0}=\t{\gamma}^{0}&=\gamma^{0}, &\m{T}^{1,q}=\t{\gamma}^{q}=i \gamma^{q}.
\end{align*}
For modified Dirac matrices
\begin{align*}
&\t{\gamma}^{a}\gamma^{b}+\t{\gamma}^{b}\gamma^{c}=I^{ab}I_{4},\\
&\t{\gamma}^{a\dagger}=\gamma^{a},
\end{align*}
where $I^{ab}=diag(1,1,1,1)$. Then, the isomorphism between the bases of $Cl_{1,3}(\mathbb{R})$ and the generators of $U(4)$ group is 
\begin{align*}
&\m{T}^{1,a}=\t{\g}^{a},\\
&\m{T}^{2,ab}=i \t{\g}^{a}\t{\g}^{b},\\
&\m{T}^{3,abc}=i \t{\g}^{a}\t{\g}^{b}\t{\g}^{c},\ \ \ \ \ \ \ \ \ \ \ \ \ \ \ \ \ (a< b < c< d).\\
&\m{T}^{4,abcd}=\t{\g}^{a}\t{\g}^{b}\t{\g}^{c}\t{\g}^{d},\\
&\m{T}^{0}=I_{4},
\end{align*}
It is easy to see that the constructed $\m{T}^{\alpha}$ satisfy 
\begin{eqnarray}
\m{T}^{\alpha\dagger}=\m{T}^{\alpha},\quad Tr\left(\m{T}^{\alpha}\m{T}^{\beta}\right)=\delta^{\alpha\beta}.
\end{eqnarray}
The commutative and anti-commutative relations of constructed $\m{T}$ are
\begin{align*}
&[\m{T}^{1,a},\m{T}^{1,b}]=-2i\m{T}^{2,ab},& &\{\m{T}^{1,a},\m{T}^{1,b}\}=0,\\
&[\m{T}^{1,a},\m{T}^{2,bc}]=0,& &\{\m{T}^{1,a},\m{T}^{2,bc}\}=2\m{T}^{3,abc},\\
&[\m{T}^{1,a},\m{T}^{2,ab}]=2i\m{T}^{1,b},& 
&\{\m{T}^{1,a},\m{T}^{2,ab}\}=0,\\
&[\m{T}^{1,a},\m{T}^{3,bcd}]=2i\m{T}^{4,abcd},& &\{\m{T}^{1,a},\m{T}^{3,bcd}\}=0,\\
&[\m{T}^{1,a},\m{T}^{3,abc}]=0,& &\{\m{T}^{1,a},\m{T}^{3,abc}\}=2\m{T}^{2,bc},\\
&[\m{T}^{1,a},\m{T}^{4,abcd}]=-2i\m{T}^{3,bcd},& &\{\m{T}^{1,a},\m{T}^{4,abcd}\}=0,\\
&[\m{T}^{2,ab},\m{T}^{2,cd}]=0,& &\{\m{T}^{2,ab},\m{T}^{2,cd}\}=-2\m{T}^{4,abcd},\\
&[\m{T}^{2,ab},\m{T}^{2,bc}]=2i\m{T}^{2,ac},& &\{\m{T}^{2,ab},\m{T}^{2,bc}\}=0,\\
&[\m{T}^{2,ab},\m{T}^{3,bcd}]=2i\m{T}^{3,acd},& &\{\m{T}^{2,ab},\m{T}^{3,bcd}\}=0,\\
&[\m{T}^{2,ab},\m{T}^{3,abc}]=0,& &\{\m{T}^{2,ab},\m{T}^{3,abc}\}=2\m{T}^{1,c},\\
&[\m{T}^{2,ab},\m{T}^{4,abcd}]=0,& &\{\m{T}^{2,ab},\m{T}^{4,abcd}\}=-2\m{T}^{2,cd},\\
&[\m{T}^{3,abc},\m{T}^{4,abcd}]=-2i\m{T}^{1,d},& &\{\m{T}^{3,abc},\m{T}^{4,abcd}\}=0.
\end{align*}
Explicitly, the constructed generators of $U(4)$ are represented by Dirac matrices
\begin{align*}
&\m{T}^{1}=\g^{0}, &
&\m{T}^{2}=i \g^{1},\\
&\m{T}^{3}=i \g^{2},&
&\m{T}^{4}=i \g^{3},\\
&\m{T}^{5}=-\g^{0}\g^{1},&
&\m{T}^{6}=-\g^{0}\g^{2},\\
&\m{T}^{7}=-\g^{0}\g^{3},&
&\m{T}^{8}=-i\g^{1}\g^{2},\\
&\m{T}^{9}=-i\g^{1}\g^{3},&
&\m{T}^{10}=-i\g^{2}\g^{3},\\
&\m{T}^{11}=-i\g^{0}\g^{1}\g^{2},&
&\m{T}^{12}=-i\g^{0}\g^{1}\g^{3},\\
&\m{T}^{13}=-i\g^{0}\g^{2}\g^{3},&
&\m{T}^{14}=\g^{1}\g^{2}\g^{3},\\
&\m{T}^{15}=-i\g^{0}\g^{1}\g^{2}\g^{3},&
&\m{T}^{0}=I_{4}.
\end{align*}
and we have
\begin{eqnarray}
[\mathcal{T}^{\alpha},\mathcal{T}^{\beta}]=f^{\alpha\beta}_{\ \ \ \gamma}\mathcal{T}^{\gamma}.
\end{eqnarray}
As an example, the Weyl representation of Dirac matrices could gives us a team of explicit matrix representation of generators of $U(4)$ group.

\section{Square root Lorentz manifold}
\subsection{Pair of entities}
We define a pair of entities
\begin{eqnarray}\label{l}
l(x)&=&i\gamma^{0}(x){\gamma }^{a}(x){{\theta }_{a}}(x) ,\\ \label{lt}
\tilde{l}(x)&=&i{\gamma }^{a}(x)\gamma^{0}(x){{\theta }_{a}}(x),
\end{eqnarray}
we call them square root metric of $(1+n)$-dimensional $\mathbf{L}$ and $\mathbf{R}$. 
This pair of entities describes square root Lorentz manifold $\mathbf{rL}$. 
Direct calculations show that the definition (\ref{l}) and (\ref{lt}) satisfy 
\begin{eqnarray}
l^{\dagger}(x)=-l(x), \quad \tilde{l}^{\dagger}(x)=-\tilde{l}(x).
\end{eqnarray} 
The Dirac matrices on $\mathbf{rL}$ has potential to write as follow
\begin{eqnarray}\nonumber
\gamma^{a\prime}(x)=\gamma^{a\prime}_{\ ij}(x) e^{\dagger\prime}_{j}(x)\otimes e^{\prime}_{i}(x)=u^{\dagger}_{i}(x)\gamma^{a\prime}(x) u_{j} (x)e^{\dagger}_{j}(x)\otimes e_{i}(x),
\end{eqnarray}
For any point $x_{0}\in \mathbf{L}$ and $\mathbf{R}$
\begin{equation}\label{ei_L}
\left.\mathbf{tr}( e^{\dagger}_{j}(x) \otimes e_{i}(x))\right|_{x\rightarrow x_{0}}=\left.e_{i}(x)e^{\dagger}_{j}(x)\right|_{x\rightarrow x_{0}}=\delta_{ij}.
\end{equation}
One simple choice of $e_{i}(x)(i=1,2\cdots, k)$ on manifold is
\begin{eqnarray} \label{bases} 
e_{1}(x)=(e^{i\theta_{1}(x)},0,0,\cdots,0), &\quad& e_{2}(x)=(0,e^{i\theta_{2}(x)},0,\cdots,0),\\
\cdots\ \ \ \ \ \ \ \ \ \ \ \ \ \ \ \ \ &\quad& e_{k}(x)=(0,0,0,\cdots,e^{i\theta_{k}(x)}).
\end{eqnarray}
The bases $\left.e^{\dagger\prime}_{i}(x)\right|_{x\rightarrow x_{0}}, (i=1,2,\cdots,k)$ on $\mathbf{rL}$ has $U(k)$ freedom to choose, locally,
\begin{eqnarray}
\left.e^{\dagger\prime}_{i}(x)\right|_{x\rightarrow x_{0}}=\left.u_{ij}(x)e^{\dagger}_{j}(x)\right|_{x\rightarrow x_{0}}, \quad \left.u(x)\right|_{x\rightarrow x_{0}}\in U(k).
\end{eqnarray}
Similarly, there is another local freedom to choose representation of components of Dirac matrices
\begin{eqnarray}\nonumber
\gamma^{a\prime}(x)=\gamma^{a\prime}_{\ ij}(x) e^{\dagger\prime}_{j}(x)\otimes e^{\prime}_{i}(x)=u^{\dagger}_{i}(x)\gamma^{a}(x) u_{j} (x)e^{\dagger\prime}_{j}(x)\otimes e^{\prime}_{i}(x),
\end{eqnarray}
with 
\begin{eqnarray}
\left.\gamma^{a\prime}_{\ ij}(x)\right|_{x\rightarrow x_{0}}= \left.u^{\dagger}_{ik}(x)\gamma^{a}_{\ kl}(x)u_{lj}(x)\right|_{x\rightarrow x_{0}}=\left.u^{\dagger}_{i}(x)\gamma^{a}(x)u_{j}(x)\right|_{x\rightarrow x_{0}}, \quad \left.u(x)\right|_{x\rightarrow x_{0}}\in U(k^{\prime}).\ \ 
\end{eqnarray}
Then, there is extra $U(k^{\prime})\times U(k)$ principal bundle on $(1+n)$-dimensional square root Lorentz manifold $\mathbf{rL}$ than Lorentz manifold $\mathbf{L}$ and
\begin{eqnarray}
k^{\prime}=k=\sqrt{2^{1+n}}.
\end{eqnarray}
Under local $U(k^{\prime})\times U(k)$ bases rotation equivalence relation, there still remains $U(k)$ physical freedom
\begin{eqnarray}
\gamma^{a}(x)=\gamma^{a}_{\ ij} e^{\dagger}_{j}(x)\otimes e_{i}(x)=\gamma^{a}_{\ ij}(x) e^{\dagger}_{j}\otimes e_{i}=\psi^{\dagger}_{i}(x)\gamma^{a} \psi_{j} (x)e^{\dagger}_{j}\otimes e_{i},
\end{eqnarray}
where
\begin{eqnarray}
\left.\psi(x)\right|_{x\rightarrow x_{0}}\in U(k)
\end{eqnarray}
isomorphic to the extra fiber space of associated bundle $UE^{*}_{1,2}(\mathbf{rL})$. 
In the language of mathematic, there are two extra $U(k)$ associated bundles $UE^{*}_{1,2}(\mathbf{rL})$ on $(1+n)$-dimensional square root metric $\mathbf{rL}$ than Lorentz manifold $\mathbf{L}$, with structure of left $U(k^{\prime})$ and right $U(k)$ action torsors
\begin{eqnarray}
UE^{*}_{1,2}(\mathbf{rL})=\frac{U(k^{\prime})\times UE^{*}_{1,2}(\mathbf{rL}) \times U(k)}{U(k^{\prime})\times U(k)},
\end{eqnarray}
where $l(x)$ and $\tilde{l}(x)$ are sections of $UE^{*}_{1}(\mathbf{rL})$ and $UE^{*}_{2}(\mathbf{rL})$ bundles, respectively.
An pair of square root metric can be written as
\begin{eqnarray}\label{lx}
l(x)&=&i\gamma^{0}_{ik}(x){\gamma }^{a}_{kj}(x)e^{\dagger}_{j}\otimes e_{i}{{\theta }_{a}}(x)=i\psi^{\dagger}_{i}(x)\gamma^{0}{\gamma }^{a}\psi_{j}(x)e^{\dagger}_{j}\otimes e_{i}{{\theta }_{a}}(x) ,\\
\tilde{l}(x)&=&i{\gamma }^{a}_{ik}(x)\gamma^{0}_{kj}(x)e^{\dagger}_{j}\otimes e_{i}{{\theta }_{a}}(x)=i\psi^{\dagger}_{i}(x){\gamma }^{a}\gamma^{0}\psi_{j}(x)e^{\dagger}_{j}\otimes e_{i}{{\theta }_{a}}(x) .
\end{eqnarray}
The total structure group of principal bundle $E^{*}(\mathbf{rL})$ on $(1+n)$-dimensional $\mathbf{rL}$ is
\begin{eqnarray}
G=U(k^{\prime})\times U(k)\times GL(1+n,\mathbb{R}), \quad k=\sqrt{2^{1+n}}.
\end{eqnarray}
The fiber space of associated bundles $UE_{1,2}^{*}(\mathbf{rL})$ isomorphic to 
\begin{eqnarray}
UE^{*}_{1,2}(U_{x_{0}})=U(k)\times GL(1+n,\mathbb{R}),
\end{eqnarray}
and has structure of $G$-torsors
\begin{eqnarray}
UE^{*}_{1,2}(\mathbf{rL})=\frac{U(k^{\prime})\times UE_{1,2}^{*}(\mathbf{rL}) \times U(k)\times GL(1+n,\mathbb{R})}{U(k^{\prime})\times U(k)\times GL(1+n,\mathbb{R})}.
\end{eqnarray}
There are two kinds of inverse metric for the pair of entities
\begin{eqnarray}\label{gl}
\bar{g}^{-1}(x)=\frac{1}{4}\mathbf{tr}[l(x)l(x)]=\frac{1}{4}\mathbf{tr}[\tilde{l}(x)\tilde{l}(x)]=-I^{ab}\theta_{a}(x)\theta_{b}(x),\\
g^{-1}(x)=\frac{1}{4}\mathbf{tr}[l(x)\tilde{l}(x)]=\frac{1}{4}\mathbf{tr}[\tilde{l}(x)l(x)]=-\eta^{ab}\theta_{a}(x)\theta_{b}(x),
\end{eqnarray}
after using $\gamma^{a\dagger}=\gamma^{0}\gamma^{a}\gamma^{0}$, where $\bar{g}^{-1}(x)$ and $g^{-1}(x)$ are inverse metric of Riemann manifold $\mathbf{R}$ and Lorentz manifold $\mathbf{L}$, respectively.
An pair of square root metric for metric of $\mathbf{R}$ and $\mathbf{L}$ are
\begin{eqnarray}\label{lxRL}
\bar{l}(x)&=&i\gamma_{0}(x){\gamma }_{a}(x){{\theta }^{a}_{\mu}}(x)dx^{\mu} ,\\ \label{tildelxRL}
\bar{\tilde{l}}(x)&=&i{\gamma }_{a}(x)\gamma_{0}(x){{\theta }^{a}_{\mu}}(x)dx^{\mu}.
\end{eqnarray}
Direct calculation gives us that the definition (\ref{lxRL}) and (\ref{tildelxRL}) satisfy 
\begin{eqnarray}
\bar{l}^{\dagger}(x)=-\bar{l}(x), \quad \bar{\tilde{l}}^{\dagger}(x)=-\bar{\tilde{l}}(x).
\end{eqnarray} 
The corresponding metric for $\mathbf{R}$ and $\mathbf{L}$ are
\begin{eqnarray}\label{gRL}
\bar{g}(x)=\frac{1}{4}\mathbf{tr}[\bar{l}(x)\bar{l}(x)]=\frac{1}{4}\mathbf{tr}[\bar{\tilde{l}}(x)\bar{\tilde{l}}(x)]=-I_{ab}\theta^{a}(x)\theta^{b}(x),\\
g(x)=\frac{1}{4}\mathbf{tr}[\bar{l}(x)\bar{\tilde{l}}(x)]=\frac{1}{4}\mathbf{tr}[\bar{\tilde{l}}(x)\bar{l}(x)]=-\eta_{ab}\theta^{a}(x)\theta^{b}(x).
\end{eqnarray}
The entities pair (\ref{lxRL}) and (\ref{tildelxRL}) corresponding principal bundle $E(\mathbf{rL})$ has total structure group
\begin{eqnarray}
\bar{G}=GL(1+n,\mathbb{R})\times U(k^{\prime})\times U(k), \quad k^{\prime}=k=\sqrt{2^{1+n}}.
\end{eqnarray}
The fiber space of associated bundle $UE(\mathbf{rL})$ isomorphic to 
\begin{eqnarray}
UE_{1,2}(U_{x_{0}})=U(k)\times GL(1+n,\mathbb{R}),
\end{eqnarray}
and has structure of $\bar{G}$-torsors
\begin{eqnarray}
UE_{1,2}(\mathbf{rL})=\frac{ GL(1+n,\mathbb{R})\times U(k^{\prime})\times UE_{1,2}(\mathbf{rL}) \times U(k)}{GL(1+n,\mathbb{R})\times U(k^{\prime})\times U(k)},
\end{eqnarray}
where $\bar{l}(x)$ and $\bar{\tilde{l}}(x)$ are sections of $UE_{1}(\mathbf{rL})$ and $UE_{2}(\mathbf{rL})$ bundles, respectively.

\subsection{Connection of extra bundles and gauge field}
The principal bundle connection $W_{\mu ij}(x)$, flavor interaction gauge field, is defined as follow
\begin{eqnarray} \label{connectionC}
\left. \nabla_{\mu}e^{\dagger}_{i}(x)\right|_{x\rightarrow x_{0}}=\left. \frac{e^{\dagger}_{i}(x)-e^{\dagger}_{i}(x_{0})}{x^{\mu}-x^{\mu}_{0}}\right|_{x\rightarrow x_{0}}=\left. \frac{(\delta_{ij}-u^{*}_{ij}(x))e^{\dagger}_{j}(x)}{\partial x^{\mu}}\right|_{x\rightarrow x_{0}}=\left. i {W}_{\mu ij}(x)e^{\dagger}_{j}(x) \right|_{x\rightarrow x_{0}} .
\end{eqnarray}
The conjugate transpose of definition (\ref{connectionC}) gives us that
\begin{eqnarray}
\nabla_{\mu}e_{i}(x)=-iW^{*}_{\mu ij}(x)e_{j}(x),
\end{eqnarray}
The covariant derivative $\nabla_{\mu}$ acting on (\ref{ei_L}) leads to
\begin{eqnarray}
W_{\mu ij}(x)=W^{*}_{\mu ji}(x).
\end{eqnarray}
The flavor interaction gauge field $W_{\mu ij}(x)$ can be expanded by generators of weak interaction gauge group $U(k)$
\begin{eqnarray} \label{gaugeA}
W_{\mu ij}(x)=W^{\alpha}_{\mu}(x)\mathcal{T}^{\alpha}_{ij}, \quad \alpha=0,1,2,\cdots,k^{2}-1.
\end{eqnarray}
In a word, the flavor interaction gauge field is defined
\begin{eqnarray}
\nabla_{\mu}e_{i}(x)=-ie_{j}(x)W_{\mu ji}(x), \quad \nabla_{\mu}e^{\dagger}_{i}(x)=iW_{\mu ij}(x)e^{\dagger}_{j}(x).
\end{eqnarray}
And the gauge fields $W_{\mu}^{\alpha}(x)$ are real valued
\begin{eqnarray}
W_{\mu}^{\alpha}(x)=W_{\mu}^{\alpha*}(x).
\end{eqnarray}
In Cartan geometry and homology theory, the differential forms are useful. Then, as we use the definition of coordinate free covariant derivative,
\begin{eqnarray}
\nabla=\nabla_{\mu}dx^{\mu},
\end{eqnarray}
it is easy to see
\begin{eqnarray}
\nabla e^{\dagger}_{i}(x)=W_{\mu ij}(x) e^{\dagger}_{i}(x) dx^{\mu}=W_{ ij}(x) e^{\dagger}_{i}(x) ,
\end{eqnarray}
where 
\begin{eqnarray}
W_{ ij}(x)=W_{\mu ij}(x) dx^{\mu} 
\end{eqnarray}
is flavor interaction gauge field connection $1$-form.

Similarly, the principal bundle connection $V_{\mu}(x)$, color interaction gauge field, is defined as follow
\begin{eqnarray}\nonumber \label{connectionD}
\left. \nabla_{\mu} [\gamma^{a}(x)]\right|_{x\rightarrow x_{0}}=\left.\frac{\gamma^{a}(x)-\gamma^{a}(x_{0})}{x^{\mu}-x^{\mu}_{0}}\right|_{x\rightarrow x_{0}}=\left.\frac{\gamma^{a}(x)-u^{\dagger}(x)\gamma^{a}(x)u(x)}{x^{\mu}-x^{\mu}_{0}}\right|_{x\rightarrow x_{0}}\\ \nonumber
= \left.\frac{[\gamma^{a}(x)-u^{\dagger}(x)\gamma^{a}(x)]+[u^{\dagger}(x)\gamma^{a}(x)-u^{\dagger}(x)\gamma^{a}(x)u(x)]}{x^{\mu}-x^{\mu}_{0}}\right|_{x\rightarrow x_{0}}\\ \nonumber
=\left.\frac{[(I_{k}-u^{\dagger}(x))\gamma^{a}(x)]+[u^{\dagger}(x)\gamma^{a}(x)(I_{k}-u(x))]}{x^{\mu}-x^{\mu}_{0}}\right|_{x\rightarrow x_{0}}\\ 
=\left. i [V_{\mu }(x)\gamma^{a}(x)-\gamma^{a}(x)\bar{V}_{\mu}(x)]\right|_{x\rightarrow x_{0}},\ \ 
\end{eqnarray}
we can see
\begin{eqnarray}
\bar{V}_{\mu}(x)=V^{\dagger}_{\mu}(x).
\end{eqnarray}
Then,
\begin{eqnarray} \label{connectionDD}
\left. \nabla_{\mu} [\gamma^{a}(x)]\right|_{x\rightarrow x_{0}} 
=\left. i [V_{\mu }(x)\gamma^{a}(x)-\gamma^{a}(x)V^{\dagger}_{\mu}(x)]\right|_{x\rightarrow x_{0}},\ \ 
\end{eqnarray}
The conjugate transpose of (\ref{connectionDD}) is
\begin{eqnarray} \label{connectionDDconjugate}
\left. \nabla_{\mu} [\gamma^{a\dagger}(x)]\right|_{x\rightarrow x_{0}} 
=\left. i [V_{\mu }(x)\gamma^{a\dagger}(x)-\gamma^{a\dagger}(x)V^{\dagger}_{\mu}(x)]\right|_{x\rightarrow x_{0}},\ \ 
\end{eqnarray}
As we have Hermitian condition on square root Lorentz manifold $\mathbf{rL}$
\begin{eqnarray}\label{HamiL}
\left.\gamma^{a\dagger}(x)\gamma^{b}(x)+\gamma^{b\dagger}(x)\gamma^{a}(x)\right|_{x\rightarrow x_{0}}=I^{ab}I_{k},
\end{eqnarray}
we act covariant derivative $\nabla_{\mu}$ on (\ref{HamiL}), after using $\gamma^{a\dagger}=\gamma^{0}\gamma^{a}\gamma^{0}$, it is easy to find out that
\begin{eqnarray}
V_{\mu}(x)=V_{\mu}^{\dagger}(x).
\end{eqnarray}
The $V_{\mu}$ is $k\times k$ matrix valued field, and can be expanded by generators of $U(k)$ group
\begin{eqnarray}
V_{\mu}(x)=V^{\alpha}_{\mu}(x)\mathcal{T}^{\alpha}, \quad \alpha=0,1,2,\cdots, k^{2}-1.
\end{eqnarray}
In a word, the color interaction gauge field $V_{\mu}(x)$ is defined
\begin{equation}\label{connectionD}
\nabla_{\mu} (\gamma^{a}(x))=i [V_{\mu }(x)\gamma^{a}(x)-\gamma^{a}(x)V_{\mu}(x)].
\end{equation}
The conjugate transpose of equation (\ref{connectionD}) is
\begin{equation}\label{connectionE}
\nabla_{\mu} (\gamma^{a\dagger}(x))=i [V_{\mu }(x)\gamma^{a\dagger}(x)-\gamma^{a\dagger}(x)V_{\mu}(x)].
\end{equation}

The connections preserving $G$ and $\bar{G}$-torsors on principal bundles $E^{*}(\mathbf{rL})$ and $E(\mathbf{rL})$ lead to the transformation rules of connections $W_{\mu ij}(x)$ and $V_{\mu}(x)$
\begin{eqnarray} \label{transforB1}
W^{\prime}_{\mu ij}(x^{\prime})=u^{*}_{ki}(x)W_{\mu kl}(x)u_{lj}(x)+u^{*}_{ki}(x) \partial_{\mu} u_{kj}(x) , \quad \left. u(x)\right|_{x\rightarrow x_{0}} \in U(k),\\
\label{transforB3}
V^{\prime}_{\mu}(x^{\prime})=u(x)V_{\mu}(x)u^{\dagger}(x)-(\partial_{ \mu} u(x)) u^{\dagger}(x), \quad \left. u(x)\right|_{x\rightarrow x_{0}} \in U(k^{\prime}), 
\end{eqnarray}
where 
\begin{eqnarray}
\left.\left.u^{*}_{ji}(x) u_{jk}(x)\right|_{x\rightarrow x_{0}}=\delta_{ik}, \quad u(x) u^{\dagger}(x)\right|_{x\rightarrow x_{0}}=I_{k}.
\end{eqnarray}
The gauge field strength tensors are defined as follows \cite{Fecko:2006zy}
\begin{eqnarray*} \label{cur2}
F_{\mu\nu ij}(x)&=&\partial_{\mu} W_{\nu ij}(x)-\partial_{\nu}W_{\mu ij}(x)-i W_{\mu ik}(x)W_{\nu kj}(x)+i W_{\nu ik}(x)W_{\mu kj}(x),\\ \label{cur3}
H_{\mu\nu}(x)&=&\partial_{\mu} V_{\nu}(x)-\partial_{\nu}V_{\mu }(x)-i V_{\mu}(x)V_{\nu}(x)+i V_{\nu}(x)V_{\mu}(x),
\end{eqnarray*}
and the transformation rules satisfy 
\begin{eqnarray}
F^{\prime}_{\mu\nu ij}(x^{\prime})= u^{*}_{ki}(x)F_{\mu \nu kl}(x)u_{lj}(x) , \quad
H^{\prime}_{\mu\nu}(x^{\prime})=u(x)H_{\mu\nu}(x)u^{\dagger}(x).
\end{eqnarray}
From the Hamiticity condition of gauge fields $W_{\mu ij}$ and $V_{\mu}$, the Hermitian condition of gauge field strengths are
\begin{eqnarray}
H_{\mu\nu}^{\dagger}(x)=H_{\mu\nu}(x), \quad
F^{*}_{\mu\nu ij}(x)=F_{\mu\nu ji}(x).
\end{eqnarray}
The gauge field strength tensors can be written by strength $2$-form
\begin{eqnarray}
H(x)=\frac{1}{2}H_{\mu\nu}(x)dx^{\mu}\wedge dx^{\nu},\quad F_{ij}(x)=\frac{1}{2}F_{\mu\nu ij}(x)dx^{\mu}\wedge dx^{\nu},
\end{eqnarray}
and
\begin{eqnarray}\label{F-form}
F_{ij}(x)&=&dW_{ij}(x)-iW_{ik}(x)\wedge W_{kj}(x),\\ \label{H-form}
H(x)&=&dV(x)-iV(x)\wedge V(x),
\end{eqnarray}
where 
\begin{eqnarray}
V(x)=V_{\mu}(x)dx^{\mu}
\end{eqnarray}
is color interaction gauge field connection $1$-form. The exterior derivative acting on (\ref{F-form}) and (\ref{H-form}) gives us Bianchi identity of strength $2$-form
\begin{eqnarray}
dH(x)- iH(x)\wedge V(x)+iV(x)\wedge H(x)&=&0,\\
dF_{ij}(x)-iF_{ik}(x)\wedge W_{kj}(x)+iW_{ik}(x)\wedge F_{kj}(x) &=&0.
\end{eqnarray}
The tensor formulation of Bianchi identity on this geometry structure as follows
\begin{eqnarray}
\partial_{[\mu}H_{\nu\rho] }(x)&=&H_{[\mu\nu}(x)V_{\rho]}(x)-V_{[\mu}(x)H_{\nu\rho]}(x),\\
\partial_{[\mu}F_{\nu\rho] ij}(x)&=&F_{[\mu\nu|ik|}(x)W_{\rho]kj}(x)-W_{[\mu |ik|}(x)F_{\nu\rho] kj}(x).
\end{eqnarray} 
\subsection{Lagrangian submanifold and Yang-Mills theory in curved space-time}
An pair of equations which satisfying the $U(k^{\prime})\times U(k)$ gauge invariant, locally Lorentz invariant and generally covariant principles are constructed in $(1+n)$-dimensional square root Lorentz manifold $\mathbf{rL}$
\begin{eqnarray}\label{fundamentalA0}
\mathbf{tr}\nabla [l(x)]= 0, \\ \label{fundamentalC}
\mathbf{tr}\nabla [\tilde{l}(x)]= 0,
\end{eqnarray}
those equations are generalized self-parallel transportation principle.
Eliminating index $x$, the explicit formulas of equations (\ref{fundamentalA0}) and (\ref{fundamentalC}) are
\begin{eqnarray*} 
\left[(i \partial_{\mu}\bar{\psi}_{i}-\bar{\psi}_{i} \tilde{V}_{\mu}+W_{\mu ij}\bar{\psi}_{j})\gamma^{a}\psi_{i} + \bar{\psi}_{i}\gamma^{a}(i \partial_{\mu} \psi_{i} +{V}_{\mu}\psi_{i}-\psi_{j} W_{\mu ji})
+i\bar{\psi}_{i}\gamma^{b}\psi_{i} \Gamma_{\ b\mu}^{a}\right]\theta_{a}^{\mu}=0,\\
\left[(i \partial_{\mu}\psi^{\dagger}_{i}-\psi^{\dagger}_{i} \tilde{V}_{\mu}+W_{\mu ij}\psi^{\dagger}_{j})\gamma^{a}\bar{\psi}^{\dagger}_{i} + \psi^{\dagger}_{i}\gamma^{a}(i \partial_{\mu} \bar{\psi}^{\dagger}_{i} +{V}_{\mu}\bar{\psi}^{\dagger}_{i}-\bar{\psi}^{\dagger}_{j} W_{\mu ji})
+i\psi^{\dagger}_{i}\gamma^{b}\bar{\psi}^{\dagger}_{i} \Gamma_{\ b\mu}^{a}\right]\theta_{a}^{\mu}=0,
\end{eqnarray*}
where
\begin{eqnarray}
\tilde{V}_{\mu}=\gamma^{0}V_{\mu}\gamma^{0} , \quad \bar{\psi}^{\dagger}=\gamma^{0}\psi.
\end{eqnarray}
The Lagrangians corresponding to equations (\ref{fundamentalA0}) and (\ref{fundamentalC}) are
\begin{eqnarray}\label{fundamentalA2}
\mathcal{L}= \bar{\psi}_{i} \gamma^{a}(i \partial_\mu \psi_{i}+V_{\mu}\psi_{i}-\psi_{j} W_{\mu ji})\theta^{\mu}_{a}+\frac{i}{2}\bar{\psi}_{i}\gamma^{b}\psi_{i} \Gamma_{\ b\mu}^{a}\theta_{a}^{\mu},\\ \label{fundamentalA3}
\tilde{\mathcal{L}}= \psi^{\dagger}_{i} \gamma^{a}(i\partial_\mu \bar{\psi}^{\dagger}_{i}+V_{\mu}\bar{\psi}^{\dagger}_{i}-\bar{\psi}^{\dagger}_{j} W_{\mu ji})\theta^{\mu}_{a}+\frac{i}{2}\psi^{\dagger}_{i}\gamma^{b}\bar{\psi}^{\dagger}_{i} \Gamma_{\ b\mu}^{a}\theta_{a}^{\mu}.
\end{eqnarray}
The last term in Lagrangian \eqref{fundamentalA2} is Yukawa coupling term $\bar{\psi}_{i}\phi \psi_{i}$ and the scalar (Higgs) field is Dirac matrix valued and originated from gravitational field
\begin{equation}\label{Higgs}
\phi=\frac{i}{2}\gamma^{b} \Gamma_{\ b\mu}^{a}\theta_{a}^{\mu}.
\end{equation}
Then, the Lagrangian \eqref{fundamentalA2} describes $U(k^{\prime})\times U(k)$ Yang-Mills theory in curved space-time. The Lagrangian (\ref{fundamentalA2}) and (\ref{fundamentalA3}) has relation with (\ref{fundamentalA0}) and (\ref{fundamentalC})
\begin{eqnarray}
\mathbf{tr}\nabla l(x)=\mathcal{L}-\mathcal{L}^{\dagger},\\
\mathbf{tr}\nabla \tilde{l}(x)=\tilde{\mathcal{L}}-\tilde{\mathcal{L}}^{\dagger}.
\end{eqnarray}
Then, we say $l(x)$ and $\tilde{l}(x)$ are a Lagrangian submanifolds in $UE^{*}_{1}(\mathbf{rL})$ and $UE^{*}_{2}(\mathbf{rL})$, respectively.
If equation (\ref{fundamentalA0}) and (\ref{fundamentalC}) being satisfied, the Lagrangian (\ref{fundamentalA2}) and (\ref{fundamentalA3}) are Hermitian
\begin{eqnarray}\label{unitary1}
\mathcal{L}=\mathcal{L}^{\dagger},\\ \label{unitary2}
\tilde{\mathcal{L}}=\tilde{\mathcal{L}}^{\dagger}.
\end{eqnarray}
So, the unitary principle (\ref{unitary1}) and (\ref{unitary2}) of quantum field theory consistent with generalized self-parallel transportation principle (\ref{fundamentalA0}) and (\ref{fundamentalC}).
The equations of motion for the Lagrangian (\ref{fundamentalA2}) and (\ref{fundamentalA3}) are
\begin{eqnarray}\label{fundamentalA1} 
\gamma^{a}(i \partial_\mu \psi_{i}+V_{\mu}\psi_{i}-\psi_{j} W_{\mu ji})\theta^{\mu}_{a}+\frac{i}{2}\gamma^{a}\psi_{i} \Gamma_{\ a\mu}^{b} \theta_{b}^{\mu} = 0, \\
\gamma^{a}(i \partial_\mu \bar{\psi}^{\dagger}_{i}+V_{\mu}\bar{\psi}^{\dagger}_{i}-\bar{\psi}^{\dagger}_{j} W_{\mu ji})\theta^{\mu}_{a}+\frac{i}{2}\gamma^{a}\bar{\psi}^{\dagger}_{i} \Gamma_{\ a\mu}^{b} \theta_{b}^{\mu} = 0,
\end{eqnarray}
and these equations conjugate transpose. Then, a pair of Lagrangians (\ref{fundamentalA2}) and (\ref{fundamentalA3}) which describes the $U(k^{\prime})\times U(k)$ Pati-Salam model type Yang-Mills theory in curved space-time are constructed.

The Yang-Mills Lagrangian for gauge bosons in this geometry can be constructed
\begin{equation}
\mathcal{L}_{Y}=\frac{-1}{2}\mathbf{tr}\left( H^{\mu\nu}H_{\mu\nu}\right)-\frac {\zeta}{2}F^{\mu\nu}_{ij}F_{\mu\nu ji},
\end{equation}
where $\zeta\in \mathbb{R}$ is constant.

In this geometry framework, the equations can be derived as follows
\begin{eqnarray} \label{relationA}
\nabla_{[\mu}\nabla_{\nu]}l &=&\frac{-1}{2}\left(\bar{\psi}_{i} \gamma^{a}\psi_{k}F_{\mu\nu kj} -F^{*}_{\mu\nu ki}\bar{\psi}_{k} \gamma^{a} \psi_{j} \right. \\ \nonumber
&+& \bar{\psi}_{i}\tilde{H}_{\mu\nu}\gamma^{a}\psi_{j} -\bar{\psi}_{i}\gamma^{a}H_{\mu\nu}\psi_{j} + \frac{i}{2}\left.\bar{\psi}_{i} \gamma^{b}\psi_{j} R^{a}_{\ b\mu\nu}\right)e^{\dagger}_{j} \otimes e_{i}\theta_{a},\\ \nabla_{[\mu}\nabla_{\nu]}\tilde{l} &=&\frac{-1}{2}\left(\bar{\psi}_{i} \gamma^{a\dagger}\psi_{k}F_{\mu\nu kj} - F^{*}_{\mu\nu ki}\bar{\psi}_{k} \gamma^{a\dagger} \psi_{j} \right. \\ \nonumber
&+& \bar{\psi}_{i}H_{\mu\nu}\gamma^{a\dagger}\psi_{j} -\bar{\psi}_{i}\gamma^{a\dagger}\tilde{H}_{\mu\nu}\psi_{j} + \frac{i}{2}\left.\bar{\psi}_{i} \gamma^{b\dagger}\psi_{j} R^{a}_{\ b\mu\nu}\right)e^{\dagger}_{j} \otimes e_{i}\theta_{a},
\end{eqnarray}
where $\tilde{H}_{\mu\nu}=\gamma^{0}H_{\mu\nu}\gamma^{0}$.
We define $ \nabla^{2}=\nabla_{[\mu}\nabla_{\nu]} dx^{\mu}\wedge dx^{\nu}$, the equation of motion of this gravity theory is constructed
\begin{equation}\label{gravity}
\mathbf{tr} \nabla^{2} [\tilde{l}(x) l(x)]=0.
\end{equation}
This equation (\ref{gravity}) is obviously $U(k^{\prime})\times U(k)$ gauge invariant, locally Lorentz invariant and generally covariant. The explicit formula of equation (\ref{gravity}) is
\begin{equation}\label{gravity0}
R=\frac{i}{4}\left(F_{abij}\psi^{\dagger}_{j}(\gamma^{a}\gamma^{b}-\gamma^{b\dagger}\gamma^{a\dagger})\psi_{i}-H_{ab}(\gamma^{a}\gamma^{b}-\gamma^{b\dagger}\gamma^{a\dagger})\right),
\end{equation}
where 
\begin{eqnarray}
{\partial_\mu}dx^{\nu}\otimes dx^{\rho} \partial_{\sigma}=\delta_{\mu}^{\nu} \delta_{\sigma}^{\rho}, \quad dx^{\mu}\otimes dx^{\nu} \partial_{\rho}\partial_{\sigma}=\delta^{\nu}_{\rho} \delta_{\sigma}^{\mu}
\end{eqnarray}
are used and 
\begin{eqnarray}
F_{abij}=F_{\mu\nu ij} \theta_{a}^{\mu}\theta_{b}^{\nu}, \quad H_{ab}=H_{\mu\nu} \theta_{a}^{\mu}\theta_{b}^{\nu}.
\end{eqnarray}
So we define a $U(k^{\prime})\times U(k)$ gauge invariant, locally Lorentz invariant, generally covariant Lagrangian
\begin{eqnarray} \label{Lg}\nonumber
\mathcal{L}_{g}=R \psi^{\dagger}_{i} \psi_{i}-i\left(F_{abij}\psi^{\dagger}_{j}(\gamma^{a}\gamma^{b}-\gamma^{b\dagger}\gamma^{a\dagger})\psi_{i}-\psi^{\dagger}_{i}H_{ab}(\gamma^{a}\gamma^{b}-\gamma^{b\dagger}\gamma^{a\dagger})\psi_{i}\right). \\ 
\end{eqnarray}
The Lagrangian (\ref{Lg}) is Hermitian
\begin{equation}
\mathcal{L}_{g}=\mathcal{L}^{\dagger}_{g}.
\end{equation}
The $R\psi^{\dagger}_{i} \psi_{i}$ in Lagrangian (\ref{Lg}) is the Einstein-Hilbert action. The equation (\ref{gravity0}) and the Einstein tensor can be derived from the Einstein-Hilbert action. 
\subsection{Conservative currents}
The Noether currents for Lagrangian system can be derived from Euler-Lagrangian equations. For action 
\begin{eqnarray}
S(\phi_{\kappa},\partial_{\mu}\phi_{\kappa} )=\int dx^{0}\wedge dx^{1}\cdots \wedge dx^{n} \bar{\mathcal{L}}(\phi_{\kappa},\partial_{\mu}\phi_{\kappa} )=\int dx^{0}\wedge dx^{1}\cdots \wedge dx^{n} \theta_{v}(x) \mathcal{L}(\phi_{\kappa},\partial_{\mu}\phi_{\kappa} ),\ \ \ 
\end{eqnarray}
The Euler-Lagrangian equations are
\begin{eqnarray}
\frac{\partial \bar{\mathcal{L}}}{\partial \phi_{\kappa}}-\partial_{\mu}\left(\frac{\partial \bar{\mathcal{L}}}{\partial(\partial_{\mu} \phi_{\kappa})}\right)=0.
\end{eqnarray}
As an example, after careful observation of the Lagrangian 
\begin{eqnarray}
\bar{\mathcal{L}}=\left[\bar{\psi}_{i} \gamma^{a}(i \partial_\mu \psi_{i}+V_{\mu}\psi_{i}-\psi_{j} W_{\mu ji})\theta^{\mu}_{a}+\frac{i}{2}\bar{\psi}_{i}\gamma^{b}\psi_{i} \Gamma_{\ b\mu}^{a}\theta_{a}^{\mu}\right] \theta_{v},
\end{eqnarray}
we set 
\begin{eqnarray}
\phi_{\kappa}=\{ \psi_{i}\},
\end{eqnarray}
then the Euler-Lagrangian equation gives us four conservative currents equations
\begin{eqnarray}
J^{\mu}_{i}=\bar{\psi}_{i}\gamma^{a}\theta_{a}^{\mu}\theta_{v},\quad \partial_{\mu}J^{\mu}_{i}=0.
\end{eqnarray}
Similarly, the conservative currents can be
\begin{eqnarray}
J^{\mu}_{i}=\bar{\psi}_{i}\gamma^{a}\theta_{a}^{\mu}\theta_{v}, \quad \gamma^{a}\psi_{i}\theta_{a}^{\mu}\theta_{v}, \quad \psi^{\dagger}_{i}\gamma^{a}\theta_{a}^{\mu}\theta_{v}, \quad \gamma^{a}\bar{\psi}^{\dagger}_{i}\theta_{a}^{\mu}\theta_{v},
\end{eqnarray}
for four Lagrangian densities $\bar{\mathcal{L}}, \bar{\mathcal{L}}^{\dagger},\bar{\tilde{\mathcal{L}}}$ and $\bar{\tilde{\mathcal{L}}}^{\dagger}$, respectively.
Note that $\Gamma^{a}_{\ b\mu}\theta_{a}^{\mu}=\left[\partial_{\mu} \theta_{b}^{\mu}+\theta_{b}^{\sigma}\Gamma^{\mu}_{ \ \sigma\mu}\right]$, we set
\begin{eqnarray}
\phi_{\kappa}=\{ \theta_{a}^{\mu}\},
\end{eqnarray}
then
\begin{eqnarray}
J^{b}=\bar{\psi}_{i}\gamma^{b}\psi_{i} \theta_{v}.
\end{eqnarray}

\section{Sheaf quantization and path integral quantization}

\subsection{Sheaf quantization}
$UE^{*}_{1}(\mathbf{rL})$ and $UE^{*}_{2}(\mathbf{rL})$ are two associated bundles on square root Lorentz manifold $\mathbf{rL}$, and $UE^{*}_{1}(\mathbf{rL})$, $UE^{*}_{2}(\mathbf{rL})$ are dual to each other
\begin{eqnarray}
\xymatrix{
UE^{*}_{1} \ar[rr]^{\gamma^{a}\rightarrow \gamma^{a\dagger}} \ar[dr]_{\pi^{*}} & &\ar[ll] UE^{*}_{2} \ar[dl]^{\pi^{*}} \\
& \mathbf{rL} &
}
\end{eqnarray}
Pair of entities $l(x)$ and $\tilde{l}(x)$ are sections of the bundles $UE^{*}_{1}(\mathbf{rL})$ and $UE^{*}_{2}(\mathbf{rL})$, respectively, 
\begin{eqnarray}
\xymatrix{
l(x) \ar[rr]^{\gamma^{a}\rightarrow\gamma^{a\dagger}} & &\ar[ll] \tilde{l}(x) \\
&\ar[ul]^{\pi^{*-1}} \mathbf{rL} \ar[ur]_{\pi^{*-1}}&
}
\end{eqnarray}
In mathematic, the sheaf space $SH_{1}(\mathbf{rL})$ and $SH_{2}(\mathbf{rL})$ are spanned by collection of one kind sections of the bundles
\begin{eqnarray}
\xymatrix{
SH_{1} \ar[rr]^{\gamma^{a}\rightarrow\gamma^{a\dagger}} & &\ar[ll] SH_{2} \\
&\ar[ul]^{\hat{\pi}^{*-1}} \mathbf{rL} \ar[ur]_{\hat{\pi}^{*-1}}&
}
\end{eqnarray}
The sheaf spaces $SH_{1}(\mathbf{rL})$ and $SH_{2}(\mathbf{rL})$ are dual to each other.
The superposition principle in quantum mechanics tells us that if the quantum state $|\Psi\rangle_{1}$ and $|\Psi\rangle_{2}$ exist, the superposition state
\begin{eqnarray}
|\Psi\rangle=\alpha_{1}|\Psi\rangle_{1}+\alpha_{2}|\Psi\rangle_{2},\quad \alpha_{1},\alpha_{2}\in \mathbb{C}, 
\end{eqnarray}
exist also. For pure state, the sheaf space $SH_{1}(\mathbf{rL})$ and $SH_{2}(\mathbf{rL})$ valued entities $\hat{l}(x)$ and $\hat{\tilde{l}}(x)$ can be defined
\begin{eqnarray}\label{lhat}
\hat{l}(x)=\sum_{\kappa}\alpha_{\kappa}(x)\alpha^{*}_{\kappa}(x)|\kappa,x\rangle \langle \kappa,x| l_{\kappa}(x),\\
\hat{\tilde{l}}(x)=\sum_{\kappa}\alpha_{\kappa}(x)\alpha^{*}_{\kappa}(x)|\kappa,x\rangle \langle \kappa,x| \tilde{l}_{\kappa}(x),
\end{eqnarray}
with the quantum field theory quantum state
\begin{eqnarray}
|\Psi(x)\rangle=\sum_{\kappa}\alpha_{\kappa}(x) |\kappa,x\rangle,
\end{eqnarray}
where $\kappa$ is sheaf space index and evaluated in an abelian group.
The orthogonalization relation of bases in sheaf spaces $SH_{1}(\mathbf{rL})$, $SH_{2}(\mathbf{rL})$ and probability complete formulas can be defined 
\begin{eqnarray} \label{orth_pure}
\langle \kappa,x|\kappa^{\prime},x^{\prime}\rangle&=&\delta(x-x^{\prime})\delta(\kappa-\kappa^{\prime}) ,\\ 
\label{pro_c_pure}
\int \mathbf{tr}\left( |\Psi(x)\rangle \langle \Psi(x)| \right)dx &=&\int \sum_{\kappa}\alpha_{\kappa}(x)\alpha^{*}_{\kappa}(x) dx=1,
\end{eqnarray}
where
\begin{eqnarray}
dx=dx^{1}\wedge \cdots \wedge dx^{n}.
\end{eqnarray}
The exterior derivative acting on (\ref{pro_c_pure})
\begin{equation}\label{lhat}
d \left[\int \mathbf{tr}\left( |\Psi(x)\rangle \langle \Psi(x)|\right) dx \right]=\int \mathbf{tr} \left[d \left(|\Psi(x)\rangle \langle \Psi(x)|\right) \right]dx=0,
\end{equation}
gives us the Sch$\ddot{o}$rdinger equations
\begin{eqnarray}\label{Schordinger}
i\frac{\partial |\Psi(x)\rangle}{\partial t}=\hat{H}(x) |\Psi(x)\rangle,\quad \hat{H}(x)=\hat{H}^{\dagger}(x),\\ \label{Schordinger0}
i\frac{\partial |\Psi(x)\rangle}{\partial x^{q}}=\hat{P}_{q}(x) |\Psi(x)\rangle,\quad \hat{P}_{q}(x)=\hat{P}_{q}^{\dagger}(x).
\end{eqnarray}

The quantum state of quantum field theory might be presented by mixed state
\begin{equation}\label{lhat}
\rho(x)=\sum_{\kappa}\eta_{\kappa}(x)|\kappa,x\rangle \langle \kappa,x|.
\end{equation}
Further, the correspongding sheaf valued entities $\hat{l}(x)$ and $\hat{\tilde{l}}(x)$ can be written
\begin{eqnarray}\label{lhat}
\hat{l}(x)=\sum_{\kappa}\eta_{\kappa}(x)|\kappa,x\rangle \langle \kappa,x| l_{\kappa}(x),\\
\hat{\tilde{l}}(x)=\sum_{\kappa}\eta_{\kappa}(x)|\kappa,x\rangle \langle \kappa,x| \tilde{l}_{\kappa}(x),
\end{eqnarray}
where $\eta_{\kappa}(x)$ are probability density of corresponding section $l_{\kappa}(x)$ and $\tilde{l}_{\kappa}(x)$.
The probability complete formulas in mixed state case is
\begin{eqnarray} 
\label{pro_c}
\int \mathbf{tr}\rho(x) dx &=&\int \sum_{\kappa}\eta_{\kappa}(x) dx=1.
\end{eqnarray}
The sheaf spaces $SH_{1}(\mathbf{rL})$ and $SH_{2}(\mathbf{rL})$ are linear spaces, which means, for example, any two entities in $SH_{1}(\mathbf{rL})$, there is a entity equals to the mixing of the two entities
\begin{eqnarray}\nonumber
&&\hat{l}(x)=\eta_{1}(x) \hat{l}_{1}(x)+\eta_{2}(x)\hat{l}_{2}(x) ; \quad \hat{l}_{1}(x), \hat{l}_{2}(x) \in SH_{1}(\mathbf{rL})\\
&\Rightarrow& \hat{l}(x)\in SH_{1}(\mathbf{rL}),
\end{eqnarray}
where 
\begin{eqnarray}
\int dx \eta_{1}(x), \int dx \eta_{2}(x)\in [0,1],
\end{eqnarray}
and 
\begin{eqnarray}
\int dx\left[ \eta_{1}(x)+\eta_{2}(x)\right]=1.
\end{eqnarray}
We call it sheaf quantization which switching study objects from single section to one kind of possible sections of the bundle. Sheaf quantization method find out a pair of linear space $SH_{1}(\mathbf{rL})$ and $SH_{2}(\mathbf{rL})$ even the $\mathbf{rL}$ is curved space-time, sheaf quantization method consistent with superposition principle.
The equations of motion for entities $\hat{l}(x)$ and $\hat{\tilde{l}}(x)$ after sheaf quantization are
\begin{equation}\label{fundamentalA00}
\mathbf{tr}\nabla [\hat{l}(x)]= 0, \quad
\mathbf{tr} \nabla^{2} [\hat{\tilde{l}}(x) \hat{l}(x)]=0.
\end{equation}
The corresponding total Lagrangian density is
\begin{eqnarray}
\hat{\mathcal{L}}=\sum_{\kappa}\eta_{\kappa}(\mathcal{L}_{\kappa}+g\mathcal{L}_{g,\kappa}+\tilde{g}\mathcal{L}_{Y,\kappa}),
\end{eqnarray}
where $g,\tilde{g}$ are Lagrange multipliers and 
\begin{eqnarray}
g,\tilde{g}\in\mathbb{R}.
\end{eqnarray} 

\subsection{The relations between sheaf quantization and path integral quantization}
The Sch$\ddot{o}$rdinger equation (\ref{Schordinger}) derives that
\begin{eqnarray}\label{path}
|\Psi(t+\Delta t, x^{q})\rangle =U(t+\Delta t,t) |\Psi(t, x^{q})\rangle = e^{-i\hat{H}(t,x^{q})\Delta t} |\Psi(t, x^{q})\rangle, 
\end{eqnarray}
where $t=x^{0}$, $q=1,2,\cdots, n$ and
\begin{eqnarray}
\Delta t\rightarrow 0.
\end{eqnarray}
By using the orthogonality relation (\ref{orth_pure}) of bases in the sheaf spaces $SH_{1}(\mathbf{rL})$ and $SH_{2}(\mathbf{rL})$, and we choose the orthogonal bases to be $|\phi_{\kappa}(t,x^{q})\rangle$ to span the quantum state of quantum field theory
\begin{eqnarray}
|\Psi(t,x^{q})\rangle=\sum_{\kappa}\alpha_{\kappa}(t,x^{q})\phi_{\kappa}(t,x^{q})|0\rangle =\sum_{\kappa}\alpha_{\kappa}(t,x^{q})|\phi_{\kappa}(t,x^{q})\rangle,
\end{eqnarray}
the equation (\ref{path}) can be written
\begin{eqnarray}\nonumber\label{path1}
\alpha_{\kappa^{\prime\prime}}(t+\Delta t, x^{q}) &=&\sum_{\kappa}\langle \phi_{\kappa^{\prime\prime}}(t+\Delta t, x^{q})|e^{-i\hat{H}(t,x^{q})\Delta t} |\phi_{\kappa}(t, x^{q})\rangle \alpha_{\kappa}(t,x^{q})\\ \nonumber
&=&\sum_{\kappa,\kappa^{\prime}}\int d\pi_{\kappa^{\prime}}(t+\Delta t,x^{q}) \langle \phi_{\kappa^{\prime\prime}}(t+\Delta t, x^{q})|\pi_{\kappa^{\prime}}(t+\Delta t,x^{q}) \rangle\\
&& \langle \pi_{\kappa^{\prime}}(t+\Delta t,x^{q})| e^{-i\hat{H}(t,x^{q})\Delta t} |\phi_{\kappa}(t, x^{q})\rangle \alpha_{\kappa}(t,x^{q}).
\end{eqnarray}
We know the relation of canonical position $|\phi_{\kappa}(t, x^{q})\rangle$ and momentum $|\pi_{\kappa}(t,x^{q}) \rangle$
\begin{eqnarray}
\langle \phi_{\kappa}(t, x^{q})|\pi_{\kappa^{\prime}}(t,x^{q}) \rangle=e^{i \pi_{\kappa}(t,x^{q})\phi_{\kappa}(t,x^{q}) }\delta_{\kappa\kappa^{\prime}},
\end{eqnarray}
this is second quantized version of 
\begin{eqnarray}
\langle x|p\rangle =e^{ipx},
\end{eqnarray}
then
\begin{eqnarray}\nonumber\label{path2}
\alpha_{\kappa}(t+\Delta t, x^{q}) &=&\sum_{\kappa}\int d\pi_{\kappa}(t+\Delta t,x^{q}) e^{i \pi_{\kappa}(t+\Delta t,x^{q})[\phi_{\kappa}(t+\Delta t,x^{q})-\phi_{\kappa}(t,x^{q})] }e^{-i\hat{H}(t,x^{q})\Delta t} \alpha_{\kappa}(t,x^{q})\\
&=&\sum_{\kappa}\int d\pi_{\kappa}(t+\Delta t,x^{q}) e^{i \pi_{\kappa}(t+\Delta t,x^{q})\dot{\phi}_{\kappa}(t,x^{q}) \Delta t }e^{-i\hat{H}(t,x^{q})\Delta t} \alpha_{\kappa}(t,x^{q})
\end{eqnarray}
There is Legendre transformation between Hamiltonian and Lagrangian
\begin{eqnarray}
\hat{L}=\sum_{\kappa}\pi_{\kappa} \dot{\phi}_{\kappa}-\hat{H}=\int dx \theta_{v} \hat{\mathcal{L}},
\end{eqnarray}
then
\begin{eqnarray}\nonumber\label{path3}
\alpha_{\kappa}(t+\Delta t, x^{q})
&=&\sum_{\kappa}\int d\pi_{\kappa}(t+\Delta t,x^{q}) e^{i( \pi_{\kappa}\dot{\phi}_{\kappa}-H)\Delta t } \alpha_{\kappa}(t,x^{q}) \\ \nonumber
&=&\sum_{\kappa}\int_{t^{\prime}=t+\Delta t} d \pi_{\kappa}(t^{\prime},x^{q}) e^{i \hat{L} \Delta t } \alpha_{\kappa}(t,x^{q})\\
&=&\sum_{\kappa}\int_{t^{\prime}=t+\Delta t} d \pi_{\kappa}(t^{\prime},x^{q}) e^{i \int \omega \hat{\mathcal{L}} } \alpha_{\kappa}(t,x^{q}),
\end{eqnarray}
where $\omega$ is volume form
\begin{eqnarray}
\omega=\theta_{v}dx^{0}\wedge dx^{1}\wedge \cdots \wedge dx^{n}
\end{eqnarray}
then
Then, the transition amplitude can be defined through path integral formula
\begin{eqnarray}
\alpha_{\kappa}(t,x^{q})=\sum_{\kappa}\int_{t^{\prime}\in (t_{0},t)} D\pi_{\kappa}(t^{\prime},x^{q}) e^{i\int \omega \hat{\mathcal{L}}[\phi_{\kappa}(t^{\prime},x^{q}),\partial_{\mu}\phi_{\kappa}(t^{\prime},x^{q}) ]} \alpha_{\kappa}(t_{0}, x^{q}). \ \
\end{eqnarray}

This section of proof shows that, the sheaf quantization method is consistent with path integral method even for quantum field theory in curved space-time. As we are using the second quantized canonical $|\phi_{\kappa}(t, x^{q})\rangle$ and momentum $|\pi_{\kappa}(t,x^{q}) \rangle$, the manifold after sheaf quantization and path integral quantization should be second quantized version symplectic manifold.

\section{Conclusion and Discussion}
\label{sec:3}
The existence of extra bundles on square root Lorentz manifold and the self-parallel transportation principle lead us to the Pati-Salam model in curved space-time and the Einstein-Cartan gravity. The relations between sheaf quantization method and path integral quantization method is proved. The prove shows that the sheaf quantization method is consistent with path integral method even the base manifold with curvature. 

The discussions about homology theory, homotopy theory, characteristic class in square root Lorentz manifold will be wonderful. The global solutions of square root Lorentz manifold with topologies $S^{1}\times S^{3}$ and $S^{1}\times S^{1}$ of base manifold are interesting. The micro support language of sheaf of square root Lorentz manifold might trigger a meaningful collision between mathematic theory and physical theory.

\section*{Acknowledgements}
We thank professor Chao-Guang Huang for long time discussions and powerful help. Without professor Huang's help, this work is impossible. We thank professor Fu-Zhong Yang for valuable discussions, I am here with deep memories. We thank professors Zhe Chang, Jun-Bao Wu, Ming Zhong, Hong-Fei Zhang, Jeffrey Zheng and Yong-Chang Huang for valuable discussions.

Data sharing is not applicable to this article as no new data were created or analyzed in this study.




%
%
%
%
%


\providecommand{\href}[2]{#2}\begingroup\raggedright\endgroup


\end{document}